\documentclass[twocolumn]{aastex62}

\usepackage{amsmath}
\usepackage{longtable}

\newcommand{\gaia}{\emph{Gaia}}
\newcommand{\hi}{H\textsc{i}}
\newcommand{\nstars}{$143$}
\newcommand{\nstarstot}{$1202$}

\graphicspath{{./}{figures/}}

\received{\today}

\begin{document}

\title{The 3D kinematics of gas in the Small Magellanic Cloud}

\correspondingauthor{C.\,E.\,M.}
\email{clairemurray56@gmail.com}

\author[0000-0002-7743-8129]{Claire E. Murray}
\altaffiliation{NSF Astronomy \& Astrophysics Postdoctoral Fellow}
\affil{Department of Physics \& Astronomy, 
Johns Hopkins University,
3400 N. Charles Street, 
Baltimore, MD 21218}

\author[0000-0003-4797-7030]{J. E. G. Peek}
\affil{Space Telescope Science Institute, 
3700 San Martin Drive, 
Baltimore, MD, 21218}

\affil{Department of Physics \& Astronomy, 
Johns Hopkins University,
3400 N. Charles Street, 
Baltimore, MD 21218}

\author[0000-0003-4019-0673]{Enrico M. Di Teodoro}
\affil{Research School of Astronomy and Astrophysics - The Australian National University, Canberra, ACT, 2611, Australia}

\author[0000-0003-2730-957X]{N.\,M. McClure-Griffiths}
\affil{Research School of Astronomy and Astrophysics - The Australian National University, Canberra, ACT, 2611, Australia}

\author[0000-0002-6300-7459]{John M. Dickey}
\affil{School of Natural Sciences, Private Bag 37, University of Tasmania, Hobart, TAS, 7001, Australia}

\author[0000-0002-9214-8613]{Helga D\'{e}nes}
\affil{ASTRON, Netherlands Institute for Radio Astronomy, Oude Hoogeveensedijk 4, 7991 PD, Dwingeloo, The Netherlands}

\begin{abstract}

We investigate the kinematics of neutral gas in the Small Magellanic Cloud (SMC) and test the hypothesis that it is rotating in a disk. To trace the 3D motions of the neutral gas distribution, we identify a sample of young, massive stars embedded within it. These are stars with radial velocity measurements from spectroscopic surveys and proper motion measurements from \emph{Gaia}, whose radial velocities match with dominant \hi\ components. We compare the observed radial and tangential velocities of these stars with predictions from the state-of-the-art rotating disk model based on high-resolution $21\rm\,cm$ observations of the SMC from the Australian Square Kilometer Array Pathfinder telescope. We find that the observed kinematics of gas-tracing stars are inconsistent with disk rotation. We conclude that the kinematics of gas in the SMC are more complex than can be inferred from the integrated radial velocity field. As a result of violent tidal interactions with the LMC, non-rotational motions are prevalent throughout the SMC, and it is likely composed of distinct sub-structures overlapping along the line of sight.
\end{abstract}


\section{Introduction} \label{sec:intro}

The Small Magellanic Cloud (SMC) is a gas-rich, low-metallicity \citep[$Z \sim 0.2\, Z_{\odot}$;][]{russell1992} dwarf irregular satellite of the Milky Way (MW). At a distance of $\sim 62$ kpc \citep[e.g.,][]{scowcroft2016}, the SMC provides an ideal laboratory for studying the structure and kinematics of dwarf galaxies at high sensitivity and resolution.

\begin{figure*}
\begin{center}
\includegraphics[width=\textwidth]{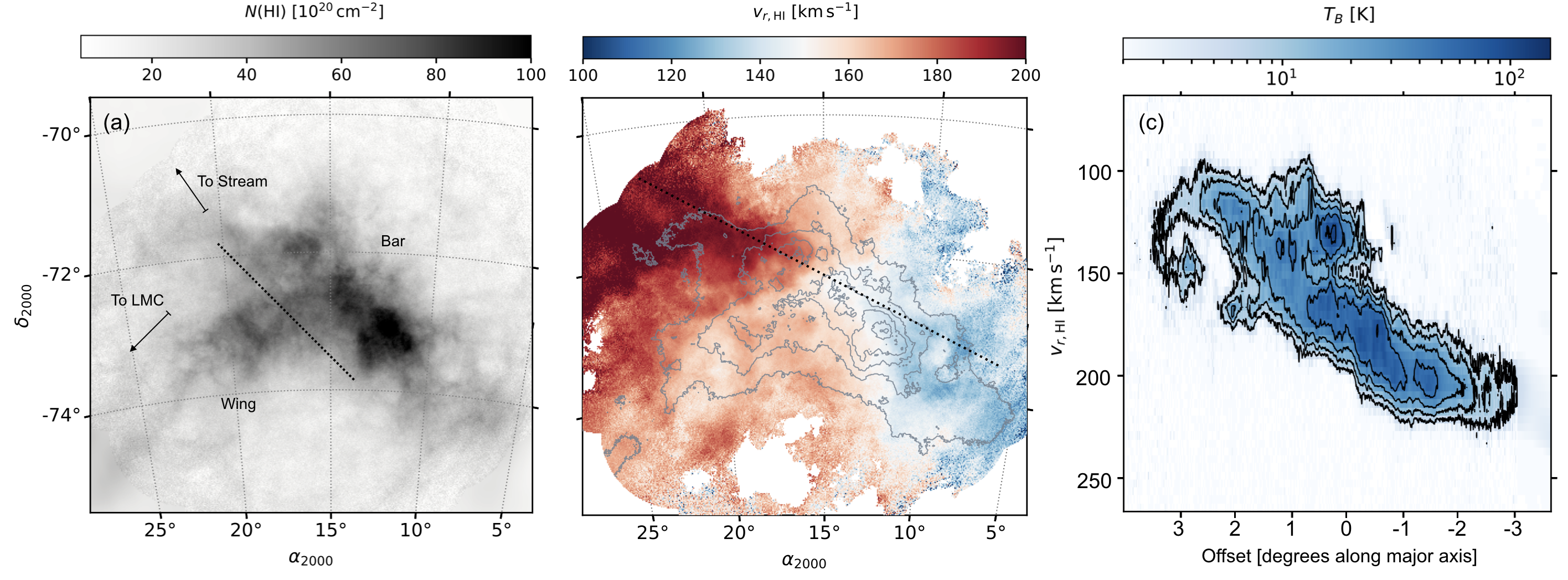}
\caption{(a) \hi\ column density map of the SMC from the ASKAP+Parkes data cube. The dotted line delineates the main ``Bar" from the ``Wing" of the SMC, and arrows indicate the directions of the Magellanic Stream and the LMC \citep[following][]{mcg2018}. (b) First moment map, masked as described in Section~\ref{sec:data_hi}. The contours (grey) display $N({\rm HI})$ at levels of $[15,35,55,75,95,115]\times10^{20}\rm\,cm^{-2}$. The dotted line (black) indicates the major axis inferred from kinematic modeling of the ASKAP+Parkes data cube \citep[center: $\alpha_{2000}=15.237^{\circ}$, $\delta_{2000}=-72.273^{\circ}$, position angle: $66^{\circ}$;][]{diteodoro2019}.
(c) Position-velocity slice through the center of the SMC and along the major axis (shown in panel b). Contours are drawn at $T_B=3\times2^n\rm\,K$.}
\label{f:smc_maps}
\end{center}
\end{figure*}

As a result of its proximity to the MW and the nearby Large Magellanic Cloud (LMC), the SMC have been strongly influenced by ongoing dynamical interactions. The SMC and LMC likely experienced a direct collision $\sim 150\rm\,Myr$ ago \citep[e.g.,][]{zivick2018} and their interactions have produced the broader, gaseous Magellanic System, including the Bridge \citep{hindman1963}, Stream \citep{mathewson1974} and Leading Arm \citep{putman1998}. In fact, low metallicities observed in the Stream and Leading Arm suggest that these structures originated in part from the SMC \citep{fox2013, donghia2016}, and this premise is consistent with most models of the Magellanic System \citep{gardiner1996, connors2006, besla2012, diaz2012}. 

The complex dynamical history of the SMC has made it challenging to understand its present-day structure and kinematics. The neutral hydrogen (\hi) in the system is highly disturbed \citep[e.g., ][]{stanimirovic1999}, featuring multi-peaked velocity profiles indicative of super giant shells \citep{hindman1967}, two separate velocity subsystems \citep[e.g.,][]{mathewson1988}, and/or hundreds of expanding shells \citep{staveleysmith1997}. However, despite this elaborate structure, the integrated velocity field of the SMC exhibits a strong gradient suggestive of a nicely rotating disk \citep{kerr1954,hindman1963,stanimirovic2004,diteodoro2019}. 

To complicate the picture further, observations of distinct stellar populations provide additional interpretations for the structure of the SMC. The oldest stars are spherically distributed within a radius of $7-12\rm\,kpc$ \citep{subramanian2012}, with little evidence for rotation from radial velocities or proper motions \citep[PMs;][]{harris2006, kallivayalil2013, gaiacollaboration2018, niederhofer2018, zivick2018}. Red clump stars and Cepheid variables are highly elongated along the line of sight \citep[$\sim20\rm\,kpc$ depth;][]{mathewson1988, nidever2013, scowcroft2016, jacyszyn2016, ripepi2017}. Furthermore, although younger stellar populations and red giant branch (RGB) stars display a velocity gradient, it appears perpendicular to the gradient seen in \hi\ \citep{evans2008, dobbie2014}. Similarly, the distance gradient observed in Cepheids from the North East to South West is $\sim90$ degrees from the minor axis implied by a rotating gas disk \citep{scowcroft2016}. 

Ultimately, although the observed distribution and kinematics of distinct mass tracers in the SMC have resulted in a diversity of interpretations for its structure, the rotating disk model derived from the \hi\ velocity field remains a fundamental benchmark for our theoretical understanding of the SMC and the broader Magellanic System. Basic properties including its center of mass, halo composition and total dynamical mass --- fundamental for numerical models --- are derived from the rotation curve analysis \citep[][]{stanimirovic2004, diteodoro2019}.
Furthermore, the SMC line of sight depth inferred from the \hi\ velocity field is routinely invoked to estimate the surface densities of molecular gas and star formation in the system. For example, although the star formation efficiency of molecular gas is the same in the SMC as within large spiral galaxies, the ratio of molecular to atomic gas is an order of magnitude lower \citep{jameson2016} implying that the SMC is ``strikingly bad" at converting its atomic gas reservoir into stars \citep{bolatto2011}. Although the interaction history between the SMC and the LMC and MW may have affected this capability by removing gas from the system,
this result is invoked to explain similarly suppressed star formation rate efficiencies of \hi-dominated galaxies at $z\sim 1-3$ \citep[e.g.,][]{wolfe2006, rafelski2016}. Although the result can be alleviated by varying uncertain disk parameters such as its inclination, a preferable model remains elusive. 

Ultimately, reconciling the disparate kinematics of gas and stars in the SMC and testing the rotating disk hypothesis requires constraints for the 3D kinematics of gas in the system. Previous models were based on the the radial velocity alone, as this is the only observable velocity component for the gas. In this paper we constrain the 3D motions of gas within the SMC using the measured 3D motions of embedded young stars (O and B type). In Section~\ref{sec:data} we present the pertinent observations and models, in Section~\ref{sec:analysis} we discuss the methods used to analyze the observations, in Section~\ref{sec:results} we present the results of comparing the observed kinematics with predictions from the rotating disk model, and in Section~\ref{sec:discussion} we discuss the implications for the SMC system. 

\section{Data}
\label{sec:data}

Although the radial motions of the SMC gas distribution are well measured by radio observations of neutral gas tracers (e.g, \hi), the transverse motions are unconstrained. We assume that massive stars whose radial velocities match with the gaseous components formed recently enough ($\sim 1-10\rm s \rm\,Myr$ lifetimes) to trace the 3D motions of the same gas that they are embedded within, and we use the measured transverse velocities of these stars to trace the transverse motions of the underlying gas. To avoid confusion by stars that have migrated away from their original gas clouds \citep{oey2018}, we make a very conservative selection to generate a small sample of \nstars\ stars whose velocities match closely with clearly defined peaks in the 21-cm spectra in their directions.  This velocity matching step rejects some $88\%$ of the original sample of stars, but the remaining stars are the most likely to show the same disk rotation that the gas shows.

\begin{figure}
\begin{center}
\includegraphics[width=0.49\textwidth]{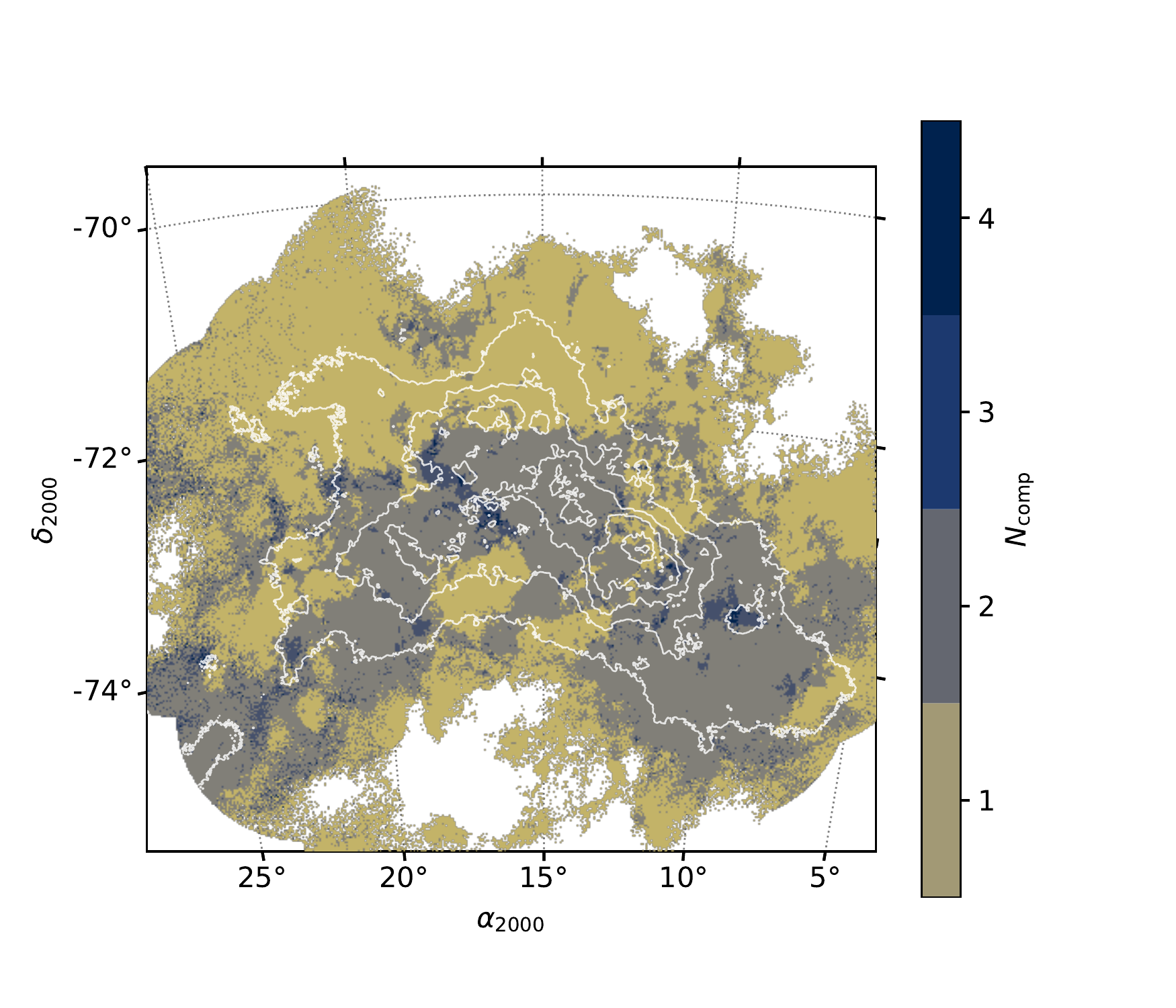}
\caption{Map of total number of significant \hi\ components ($N_{\rm comp}$), masked as described in Section~\ref{sec:data_hi}. White contours denote $N({\rm HI})$ at levels of $[15,35,55,75,95,115]\times10^{20}\rm\,cm^{-2}$.}
\label{f:comp_count}
\end{center}
\end{figure}

\subsection{Neutral gas}
\label{sec:data_hi}

As the SMC is rich in neutral atomic gas \citep[e.g.,][]{leroy2007}, we use \hi\ to trace the gas distribution. New observations of the SMC at $21\rm\,cm$ were recently obtained by the Australian Square Kilometer Array Pathfinder \citep[ASKAP;][]{deboer2009} during Commissioning and Early Science observations in November 2017. These data were combined with single-dish observations of the SMC from the Parkes Telescope as part of the HI4Pi survey \citep{hi4pi2016}. For details concerning the observations and data reduction for the ASKAP+Parkes cube, we refer the reader to \citet{mcg2018}. The final data cube we use has a root mean square (RMS) noise in $21\rm\,cm$ brightness temperature ($T_B$) of $\sigma_{T_B} = 0.7\rm\,K$, a pixel size of $7^{\prime \prime}$ ($35^{\prime \prime} \times 27^{\prime \prime}$ angular resolution), and velocity channel width of $3.9\rm\,km\,s^{-1}$ \citep{mcg2018, diteodoro2019}. For comparison with stellar velocities, we convert the \hi\ velocities from the Local Standard of Rest to the Heliocentric reference frame. Hereafter, all quoted velocities will be in the Heliocentric reference frame.

In Figure~\ref{f:smc_maps} we display a map of the \hi\ column density ($N({\rm HI})$) of the SMC from the ASKAP+Parkes data cube (panel a), the first moment map (panel b), and a position-velocity slice along the major axis inferred from kinematic modeling of the \hi\ data cube by \citet{diteodoro2019} (panel c, see Section~\ref{sec:kinmodel}). The moment map is masked as in \citet{diteodoro2019}, wherein a flood-fill algorithm applied to a smoothed version of the data cube (smoothed with Gaussian kernel of FWHM $70''$) identified pixels with emission brighter than $\sim10\sigma_{\rm HI}$, out to pixels with $\sim4\sigma_{\rm HI}$ (where $\sigma_{\rm HI}$ is the RMS of the smoothed data).

\subsection{Stars}

We select a sub-sample of SMC stars with measured radial velocities from the spectroscopic survey of the SMC by the 2-degree-field instrument at the Anglo-Australian Telescope \citep[2045 stars;][]{evans2008} and from the Runaways and Isolated O-Type Star Spectroscopic Survey of the SMC (RIOTS4), a uniformly selected survey of young field stars in the SMC using the Inamori-Magellan Areal Camera and Spectrograph (IMACS) and Magellan Inamori Kyocera Echelle (MIKE) on the Magellan telescopes \citep[374 stars;][]{lamb2016}. Of this sample, 1484 stars are of spectral type O or B and have significant radial velocity ($v_r$) measurements (defined as $v_{r}/\sigma_{v_r}>5$, where $\sigma_{v_r}$ is the uncertainty on the radial velocity measurement).

\begin{figure*}
\begin{center}
\includegraphics[width=0.97\textwidth]{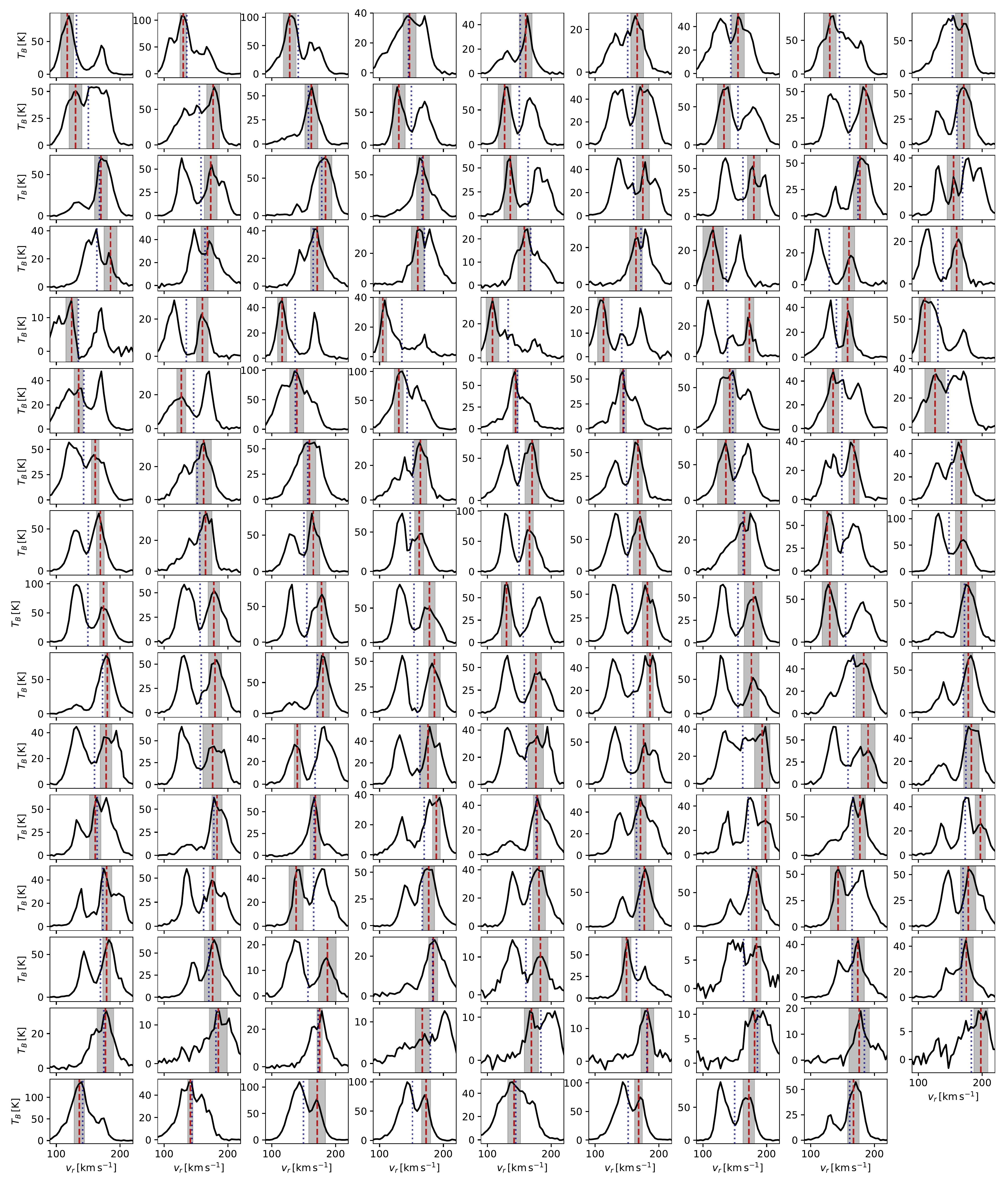}
\caption{Results of matching observed radial velocities of young, massive stars (red dashed) with the radial velocities of distinct \hi\ peaks along the same LOS (black). For each of the \nstarstot\ stars with radial and transverse velocity measurements, over 1000 trials we sample random velocities within $\pm1\sigma_{v}$ (grey) and find the \nstars\ LOS which ``match" (see Section~\ref{sec:matches}). The $\pm1\sigma$ uncertainties on the $v_{r,\rm star}$ are illustrated (grey shading), and the first moment (Figure~\ref{f:smc_maps}b) is included (blue dotted).}
\label{f:ob_matched}
\end{center}
\end{figure*}

To determine the transverse velocities of the selected stars, we cross-match them with SMC stars from the second data release (DR2) of the \gaia\ mission \citep{gaiacollaboration2018}. The targets observed by \citet{evans2008} and \citet{lamb2016} are selected from the original catalog by \citet{massey2002}, who reported a positional uncertainty of $1''$ per target. We begin with a conservative search radius of $3\sigma$ and find unique 1451 \emph{Gaia} sources within $3''$ of the selected stars. All matched targets have significant PM ($\mu$) detections in right ascension ($\alpha$) and declination ($\delta$) ($\mu_{\alpha}/\delta_{\mu_{\alpha}} > 5$ and  $\mu_{\delta}/\delta_{\mu_{\delta}} > 5$ where $\delta_{\mu_{\alpha}}$ and $\delta_{\mu_{\delta}}$ are the uncertainties in RA and Dec proper motions respectively). The matched stars represent the youngest, brightest end of the population observed by \emph{Gaia}; for example, the median \emph{Gaia} $G$ magnitude for the matched sample is $G=15.3$, compared with the median magnitude for the full SMC sample of $G=19$. We further refine the sample by comparing the observed flux densities, and remove $249$ objects with discrepant $G-B$ colors (defined as being further from the mean $G-B$ color of $0.16$ by more than $\pm1\sigma = 0.47$).  The final sample contains \nstarstot\ sources. 

\subsection{Systemic proper motion}

To estimate the PMs due to internal motions of the SMC, we subtract the systemic motion of the galaxy from the measured \emph{Gaia} PMs. For consistent comparison with the latest dynamical modeling of the system (see Section~\ref{sec:kinmodel}), we adopt the measured center of mass (COM) PM derived from HST photometry by \citet{kallivayalil2013}, who estimated $\mu_{W,0} = -0.772\pm0.063 \,\rm mas\,yr^{-1}$ and $\mu_{N,0}  = -1.117\pm0.061\,\rm mas\,yr^{-1}$. We note that many studies have estimated the COM PM of the SMC \citep[e.g.,][]{piatek2008, kallivayalil2013, vandermarel2016, zivick2018, niederhofer2018}, and all estimates are largely consistent within uncertainties. 

\subsection{Dynamical model}
\label{sec:kinmodel} 

For comparison with the observed kinematics, we use the results of a recent dynamical model in the SMC by \citet[][hereafter dT19]{diteodoro2019} based on the ASKAP+Parkes \hi\ data cube. dT19 modeled the SMC as a disk undergoing circular rotation. Using a Markov Chain Monte Carlo (MCMC) approach, they fitted the observed velocity field (e.g., Figure~\ref{f:smc_maps}b) to derive global properties of the system, including kinematic center, position and inclination angles, transverse and systemic velocities, and then applied a tilted-ring analysis to decompose the velocity field and extract the rotation velocity as a function of radius. 

\begin{figure*}
\begin{center}
\includegraphics[width=\textwidth]{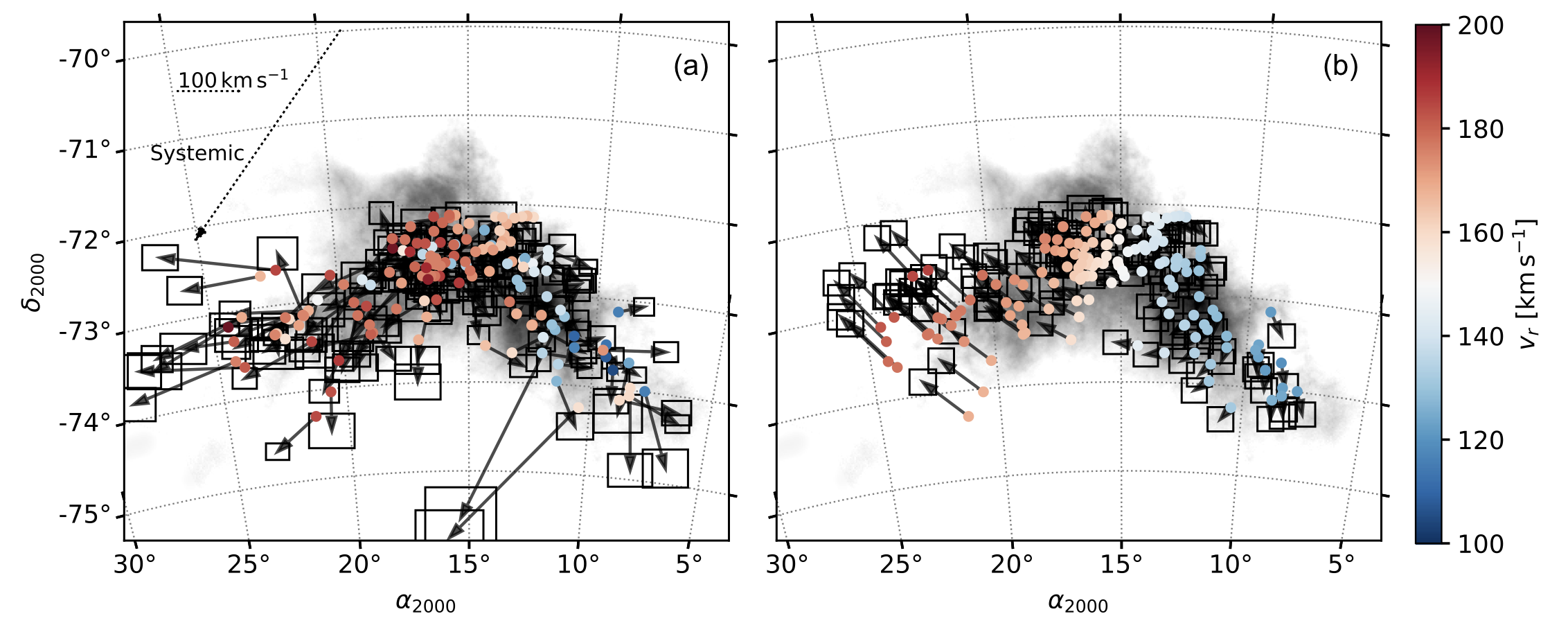}
\caption{(a): Radial velocities (color) and residual PMs (vectors) for the sample of \nstars\ stars with significantly-detected radial and transverse motions from spectroscopy \citep{evans2008, lamb2016} and \emph{Gaia} \citep{gaiacollaboration2018}, whose radial velocities match significant \hi\ components along the line of sight (c.f. Figure~\ref{f:ob_matched}). All vectors are constructed to illustrate PMs over the next $7.2\rm\,Myr$ \citep[arbitrarily, as in][]{vandermarel2016}. The systemic PM of the SMC \citep{kallivayalil2013} is included at top left (dotted), as is a scale bar to illustrate $100\rm\, km\,s^{-1}$ (dashed). (b) Same, for the predicted radial velocities and residual PMs (black vectors) from the SMC rotation model by dT19.} 
\label{f:proper_motions}
\end{center}
\end{figure*}

Using the dT19 results, we predict the velocities for stars within a rotating SMC based on their coordinates. 
To compute each velocity component, we follow \citet[][hereafter vdM2001]{vandermarel2001} and \citet[][hereafter vdM2002]{vandermarel2002}. We assume the same global SMC parameters as derived in the dT19 model, including: the SMC COM ($\alpha_{0, 2000}=15.237 \pm 0.380^{\circ}, \,\delta_{0, 2000}=-72.273 \pm 0.290^{\circ}$), distance ($D_0=63\pm5\rm\,kpc$), COM PM  \citep[$\mu_{W,0} =-0.772\pm0.063,\, \mu_{N,0} = -1.117\pm 0.061\,\rm mas\,yr^{-1}$;][]{kallivayalil2013}, systemic LOS velocity (Heliocentric, $v_{r,\rm sys}=157\pm2\rm\,km\,s^{-1}$), inclination ($i=55\pm9^{\circ}$), position angle of the line of nodes ($\rm PA=66\pm8^{\circ}$; measured counter-clockwise from North) and precession and nutation rate ($\partial i/\partial t = -281\pm100\rm\,^{\circ}/Gyr$). From the assumed COM PM, we also compute the transverse velocity of the COM and its angle on the sky. 

Given these global parameters, for each star, we compute its angular coordinates on the sky (Equations 1-3; vdM2001), its radius in the galaxy frame (Equation 9; vdM2002), and its ensuing predicted rotation velocity from the dT19 model. We then follow vdM2002 to compute the individual velocity contributions from the SMC COM (Equation 13), its precession and nutation (Equation 16) and internal motions (Equation 21). The sum of these components comprise the predicted velocities in the frame of the SMC. Finally, we extract the predicted radial velocity ($v_{r, \rm Model}$), and the PMs to the West and North ($\mu_{W,\rm Model},\, \mu_{N,\rm Model}$; Equation 9; vdM2002) for comparison with the observed sample. As an estimate of the uncertainties in these predictions, we propagate the quoted uncertainties in the nine global SMC parameters from dT19 through the analysis.

\section{Analysis}
\label{sec:analysis}

\subsection{Line of Sight Complexity}
\label{sec:complexity}

Although the radial velocity structure of \hi\ in the SMC is complex, it has been shown that a significant fraction of OB stars have coincident radial velocities with \hi\ structures along the same LOS, and are therefore embedded within the \hi\ distribution \citep{lamb2016}. Although the majority of young stars are associated with some \hi\ emission, our goal is to construct a pure sample of targets which are most likely to trace the motions of dominant \hi\ components.

We begin by identifying significant \hi\ components along each line of sight in the ASKAP+Parkes data cube via the derivative of $T_B(v_r)$. First, we re-sample each $T_B(v_r)$ spectrum to $0.5\rm\,km\,s^{-1}$ resolution and smooth with a Gaussian kernel of width $8$ channels to suppress the influence of noise on selecting spurious peaks. We then compute the first numerical derivative of each spectrum and identify components as local maxima with significant emission ($T_B > 5\rm\,K$). Figure~\ref{f:comp_count} displays the total number of significant \hi\ components along each LOS. Although the outskirts of the SMC typically feature a single strong \hi\ component, within the main body of the SMC the radial velocity structure of the gas is more complex and there are at least two \hi\ components. This is evident in Figure~\ref{f:smc_maps}c: in addition to the ``main disk" of the SMC along the major axis inferred by rotation models, there are several significant velocity components evident in the position-velocity diagram.

\subsection{Matching radial velocities}
\label{sec:matches}

To identify which stars are moving with the identified \hi\ components, we compare the observed stellar radial velocities ($v_{r,\rm star}$) with the radial velocities of the  \hi\ components ($v_{r, \rm HI}$). To select the most likely sources to trace the dominant \hi\ along each line of sight, we design a method for matching stars with \hi\ components which accounts for the uncertainties in $v_{r,\rm star}$, which can be large \citep[$\sigma_{v_{r,star}}\sim5-30\rm\, km\,s^{-1}$;][]{evans2008,lamb2016}.

For each star, we select $N=1000$ velocities distributed normally with mean $v_{r,\rm star}$ and $\sigma=\sigma_{v_{r,\rm star}}$. For each \hi\ component, we construct similar normally-distributed velocity distributions with mean $v_{r,\rm HI}$ and an assumed $\sigma = \sigma_{\rm HI} \rm\,km\,s^{-1}$. We then use a two-sample Anderson-Darling test \citep[A-D;][]{ad1954} to test for the null hypothesis that the stellar and \hi\ velocity distributions are drawn from the same parent population. A match is achieved if the null hypothesis cannot be rejected with high confidence ($>99\%$), and if the brightness temperature of the \hi\ peak is above a certain fraction of the max $T_B$ along the line of sight ($r_{\rm peak}$, where $r_{\rm peak} = T_{B,\rm peak}/\max({T_{B}(v_r)})
$), to ensure that matches are with dominant \hi\ peaks.

\begin{figure*}
\begin{center}
\includegraphics[width=\textwidth]{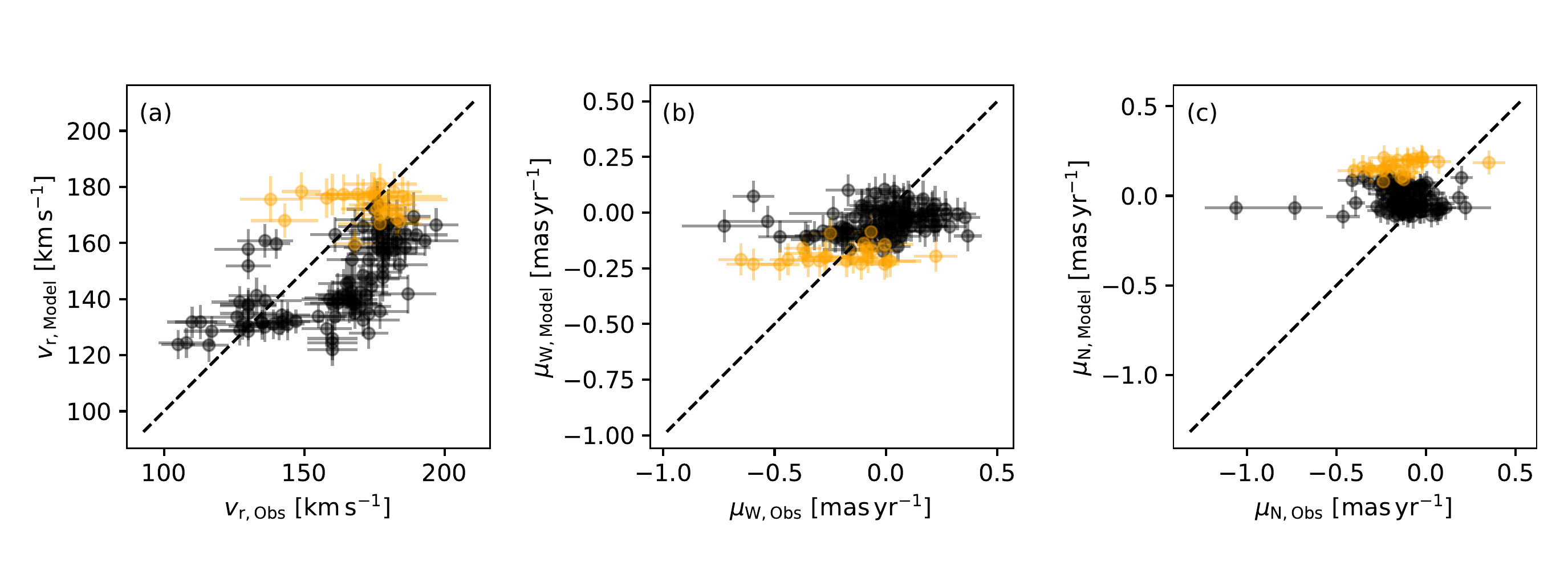}
\caption{Comparison between radial velocity ($v_{\rm r}$) and PMs to the West and North ($\mu_W$ and $\mu_N$) observed (``Obs") and predicted by dT19 (``Model") for the sample of \nstars{} stars whose radial velocities match significant \hi\ components. The error bars on $v_{r,\rm Obs}$ are $1\sigma$ uncertainties \citep{evans2008, lamb2016}, and the error bars on $\mu_{W, \rm obs}$ and $\mu_{N,\rm obs}$ are $1\sigma$ uncertainties from the \emph{Gaia} DR2 catalog \citep{gaiacollaboration2018}. The error bars for the ``Model" parameters are estimated by propagating the uncertainties in the global SMC parameters through the velocity calculations (see Section~\ref{sec:kinmodel}). We divide the sample by location in the SMC (see Figure~\ref{f:smc_maps}a for definitions): the Bar (black), and Wing (orange).}
\label{f:smc_model_scatter}
\end{center}
\end{figure*}

To select the optimal values for the matching parameters ($\sigma_{\rm HI}$ and $r_{\rm peak}$), we repeat the match process for a range of values ($1<\sigma_{\rm HI}<10\rm\,km\,s^{-1}, \, 0<r_{\rm peak}<1$), and select the values which maximize the number of matches while minimizing the contamination fraction. To quantify the contamination of the sample due to random interlopers with coincidental radial velocities to the \hi\ peaks, we repeat the matching process using simulated samples of \nstarstot\ stars with random velocities and uncertainties drawn from the observed distributions over 1000 trials. The estimated number of spurious matches is computed as the median over all trials. We find that the choice of $\sigma_{\rm HI}$ and $r_{\rm peak}$ does not have a significant effect on the results, and the contamination fraction is $\sim 25\%$ for all tested values. We will discuss the effect of this contamination fraction in Section~\ref{sec:results}. We select $\sigma_{\rm HI}=5\rm\, km\,s^{-1}$ and $r_{\rm peak}=0.3$.

We note that this approach produces statistically indistinguishable results from other methods of matching $v_{r,\rm star}$ and $v_{\rm HI}$, including: (1) by-eye matching, (2) applying a range of absolute cuts to $\Delta v_r = |v_{r, \rm star} - v_{r,\rm HI}|$, and (3) computing the area of overlap between the $v_{r,\rm star}$ and $v_{r, \rm HI}$ distributions and applying a sigma cut. 

To determine the final sample, we repeat the matching process with the selected match criteria over an additional 1000 trials, and select all stars which are successful matches in more than $95\%$ of trials. The result is a sample of \nstars\ stars. The matched sources are shown in Figure~\ref{f:ob_matched}. In the Appendix, we include similar figures for all LOS (Figures~\ref{f:match_summary0} through~\ref{f:match_summary8}).

Although the final sample represents only $\sim 12\%$ of the initial sample of stars with both radial and transverse velocity constraints (\nstarstot\ sources), this does not mean that the remaining $88\%$ of the sample are not associated with \hi\ gas. In fact, we observe that the majority of the initial sample of \nstarstot\ stars ($\sim80\%$) is associated with significant \hi\ emission, defined has there being significant \hi\ emission at the radial velcoity of the star ($T_B(v_{r, \rm star})>5\rm\,K$). This indicates that most stars are embedded somewhere within the neutral gas distribution, if not associated with the strongest peeaks. The final sample of \nstars\ matched stars represents our highest confidence for tracing the \hi\ kinematics with minimal contamination. We further observe that although relaxing the matching criteria (e.g., including wider uncertainty ranges, selecting different brightness temperature cuts to define \hi\ components, etc) changes the number and selection of stars which are included in the matched sample, it does not significantly affect the statistics of the sample (e.g., estimated contamination fraction) or the ensuing conclusions. 

\section{Results}
\label{sec:results}

In Figure~\ref{f:proper_motions}a we display the residual PMs (i.e., with the COM PM from \citet{kallivayalil2013} subtracted) for the \nstars{} stars whose radial velocities match with significant \hi\ peaks. The stars exhibit a radial velocity gradient in agreement with the \hi\ radial velocity field (Figure~\ref{f:smc_maps}b). However, the observed residual PMs do not exhibit a clear signature of rotation. For comparison, in Figure~\ref{f:proper_motions}b, we display the radial velocities and residual PMs predicted by the dT19 rotation model. As expected, the modeled residual PMs exhibit a clear signature of rotation.

Comparing the panels of Figure~\ref{f:proper_motions}, although the observed and modeled kinematics are not consistent throughout the SMC, the level of agreement may depend on location within the system. For example, a subset of stars have large observed tangential velocities ($\sim 100\rm\,km\,s^{-1}$). These large velocities are likely dominated by dynamical effects \citep{oey2018}. Stars in the the Wing region (Southeast, c.f. Figure~\ref{f:smc_maps}a) are moving towards the LMC via the Magellanic Bridge as a result of their recent interaction history \citep{zivick2019}. Furthermore, this net PM of the Wing relative to the Bar indicates that these two regions may be kinematically distinct \citep{oey2018}. Although stars and gas throughout the SMC are affected at some level by the same dynamical interactions, in subsequent analysis we will distinguish stars in the Wing and Bar to account for the possibility that stars in different regions are tracing distinct populations. 

In Figure~\ref{f:smc_model_scatter} we compare the observed and predicted radial (LOS) velocity (a), and the residual PMs to the West and North (b and c) for the \nstars{} matched stars. Stars located in Wing and Bar are distinguished from each other for clarity (see Figure~\ref{f:smc_maps}a for definitions). 

Figures~\ref{f:proper_motions} and \ref{f:smc_model_scatter} indicate that the observed kinematics are largely inconsistent with the predictions from the rotating disk model. To quantify the agreement, we use the K-sample A-D statistic. First, we bootstrap the sample with replacement over 1000 trials, and within each trial use a simple Monte Carlo exercise to draw random values of each velocity component within $\pm3\sigma$ to compute the A-D statistic. The final value is a median over all trials. We find that the null hypothesis that the observed and modeled velocity distributions are drawn from the same parent population is ruled out with $>99\%$ confidence for $v_r$, $\mu_W$ and $\mu_N$ for the full sample. When treating the two regions of the SMC separately, the agreement does not improve significantly: the null hypothesis is ruled out in all cases with $>99\%$ confidence again, with the exception of the radial velocity of the Wing ($98\%$ confidence). To explicitly and conservatively account for the estimated $25\%$ contamination from random interlopers, we repeat the above exercise and exclude the worst $25\%$ in each trial (defined as those points with the largest absolute difference between observed and modeled velocities). We find consistent results as for the full sample (agreement ruled out with $>99\%$ confidence, despite some improvement in the Bar and Wing sub-samples) except for the cases of $\mu_W$ for the full sample ($98\%$ confidence). 

\section{Discussion}
\label{sec:discussion}

Overall, the observed 3D kinematics of \hi\ in the SMC traced by young massive stars are inconsistent with predictions from the rotating disk model. In the following, we discuss how to reconcile this result within the context of the SMC.  

First, it is important to consider alternatives to the hypothesis that the SMC is well-represented by a single rotating disk. As discussed by dT19, there are several caveats to the disk model, including the fact that the disk is not razor-thin (as must be assumed to simplify the velocity component computation), the rotation is not necessarily perfectly circular and the disk is not necessarily axisymmetric. In addition to these caveats, there is clear evidence that significant components of the system (e.g., the Wing) are dominated by radial motions due to the influence of the LMC \citep[e.g., ][]{zivick2019}. The fact that the rotation model performs as well as it does at reproducing the radial velocity field of the gas suggests that the large scale motion is dominated by rotation, and the discrepant stellar kinematics observed here may reflect only local, non-circular motions. Although a subset of our sample likely does trace local motions (e.g., effects of stellar feedback on small scales), the majority were selected to trace the dominant \hi\ components. As seen in Figure~\ref{f:smc_model_scatter}a, the observed radial velocities of stars matched to \hi\ peaks trace distinct, coherent components in radial velocity, one at $\sim 130\rm\,km\,s^{-1}$ and one at $\sim160\rm\,km\,s^{-1}$, which are found in both the Wing and the Bar. The intensity-weighted mean velocity along each LOS used to  predict 3D gas motions can be insensitive to this multiplicity, especially if there are multiple components with similar intensities. To illustrate this effect, in Figure~\ref{f:ob_matched}, we include the first moment for comparison with the \hi\ peak velocities along each LOS. In addition, we note that although \emph{Gaia} PMs in the SMC display a marginal signal of rotation \citep[][Figure 16]{gaiacollaboration2018}, it is inconsistent with the rotation inferred from the \hi.

An alternative explanation for the discrepancy between the observed and model kinematics is that the SMC is not a single system, but is rather composed of multiple structures along the LOS. As illustrated in Figure~\ref{f:comp_count} and Figure~\ref{f:smc_model_scatter}, the structure of the SMC is complex, and exhibits multiple, significant \hi\ components separated in radial velocity. This velocity structure has previously been attributed to separate sub-systems within the SMC \citep[e.g., the Mini Magellanic Cloud and the SMC Remnant;][]{mathewson1984}, which may be separated in distance along the LOS by $>10\rm\,kpc$ \citep{mathewson1986}. Based on measurements of UV absorption lines towards stars in the SMC, the lower-velocity \hi\ component appears to be sitting in front of the higher-velocity \hi\ component along the LOS \citep{songaila1986, wayte1990, danforth2002, welty2012}. This implies that the SMC is being ripped apart by its interactions with the LMC, in agreement with tidal models of the system \citep{murai1980}, resulting in disparate remnant structures. 

If the SMC is composed of separate sub-systems at different distances, it is possible that the apparent rotation seen in the radial velocity field of the gas is dominated by rotation within one component over the other. However, exactly how we can reconcile potential rotation in this component alone with the non-rotating kinematics of the rest of the SMC is uncertain, and not incorporated into current theoretical models. In addition, it has been shown that models with purely radial motions can reproduce observed velocity fields just as well as disk models with purely rotational motion \citep[e.g.,][]{labini2019}. Furthermore, the derived global properties of the system, including its dynamical center and mass --- crucial input parameters to numerical models of the broader Magellanic System \citep{besla2012} --- are based on fits to the global radial velocity field (e.g., including all velocities along each LOS; Figure~\ref{f:smc_maps}). If only subsets of the SMC are actually rotating, these models (and the resulting parameters) would still need to be revisited. 

\section{Summary and Conclusions}

In this paper, we present new constraints for the 3D kinematics of gas in the SMC. We trace the transverse motions of \hi\ in the system for the first time using the observed proper motions of massive, young stars whose radial velocities match with significant \hi\ components identified from new observations by the ASKAP telescope \citep{mcg2018}. We test the hypothesis that gas in the SMC is rotating in a disk by comparing the observed radial and transverse velocities of young stars with predictions from the latest model of the radial velocity field from dT19. Our main conclusions are the following:

\begin{itemize}
    \item The observed motions of gas-tracing stars are inconsistent with the predictions from the dT19 rotation model, even when the contamination from random interlopers is explicitly taken into account (Figures~\ref{f:proper_motions},~\ref{f:smc_model_scatter}). 
    \item The 3D kinematics of the SMC are more complex than can be accurately inferred from the integrated radial velocity field alone. 
    \item Due to violent tidal and ram-pressure interactions with the LMC and the MW halo, it is likely that the SMC is composed of overlapping sub-structures, whose properties should be considered in future models of the system for extracting basic parameters such as the COM, dynamical mass and depth along the LOS. 
\end{itemize}

Overall, this study represents one step in the process of unraveling the structure and kinematics of the SMC. In the near future, with the third data release from \emph{Gaia}, we will have access to radial velocity measurements towards a new sample of stars throughout the SMC with which we will further dissect the kinematics of individual gaseous components. In addition, with resolved stellar photometry, we will constrain the distances to individual dusty components and complete the full 6D phase space of the system (3D positions, 3D velocities). As the Magellanic Clouds are the only systems for which we can achieve this level of resolution, this will not only inform our understanding of these systems, but also our understanding of dwarf galaxy interactions throughout cosmic time.

\acknowledgements{
C.E.M. is supported by an NSF Astronomy and Astrophysics Postdoctoral Fellowship under award AST-1801471.
N.Mc.-G. acknowledges funding from the Australian Research Council via grant FT150100024. 
This work took part under the program Milky-Way-Gaia of the PSI2 project funded by the IDEX Paris-Saclay, ANR-11-IDEX-0003-02.
This research has made use of NASA's Astrophysics Data System.
The Australian SKA Pathfinder is part of the Australia Telescope National Facility which is managed by CSIRO. Operation of ASKAP is funded by the Australian Government with support from the National Collaborative Research Infrastructure Strategy. ASKAP uses the resources of the Pawsey Supercomputing Centre. Establishment of ASKAP, the Murchison Radio-astronomy Observatory and the Pawsey Supercomputing Centre are initiatives of the Australian Government, with support from the Government of Western Australia and the Science and Industry Endowment Fund. We acknowledge the Wajarri Yamatji people as the traditional owners of the Observatory site. }

\software{Astropy \citep{astropy2013}, NumPy \citep{vanderwalt2011}, matplotlib \citep{hunter2007}, glue \citep{beaumont2015}}

\bibliography{ms}

\appendix
\section{Matching radial velocities}
\label{a:match}

\begin{figure*}
\begin{center}
\includegraphics[width=\textwidth]{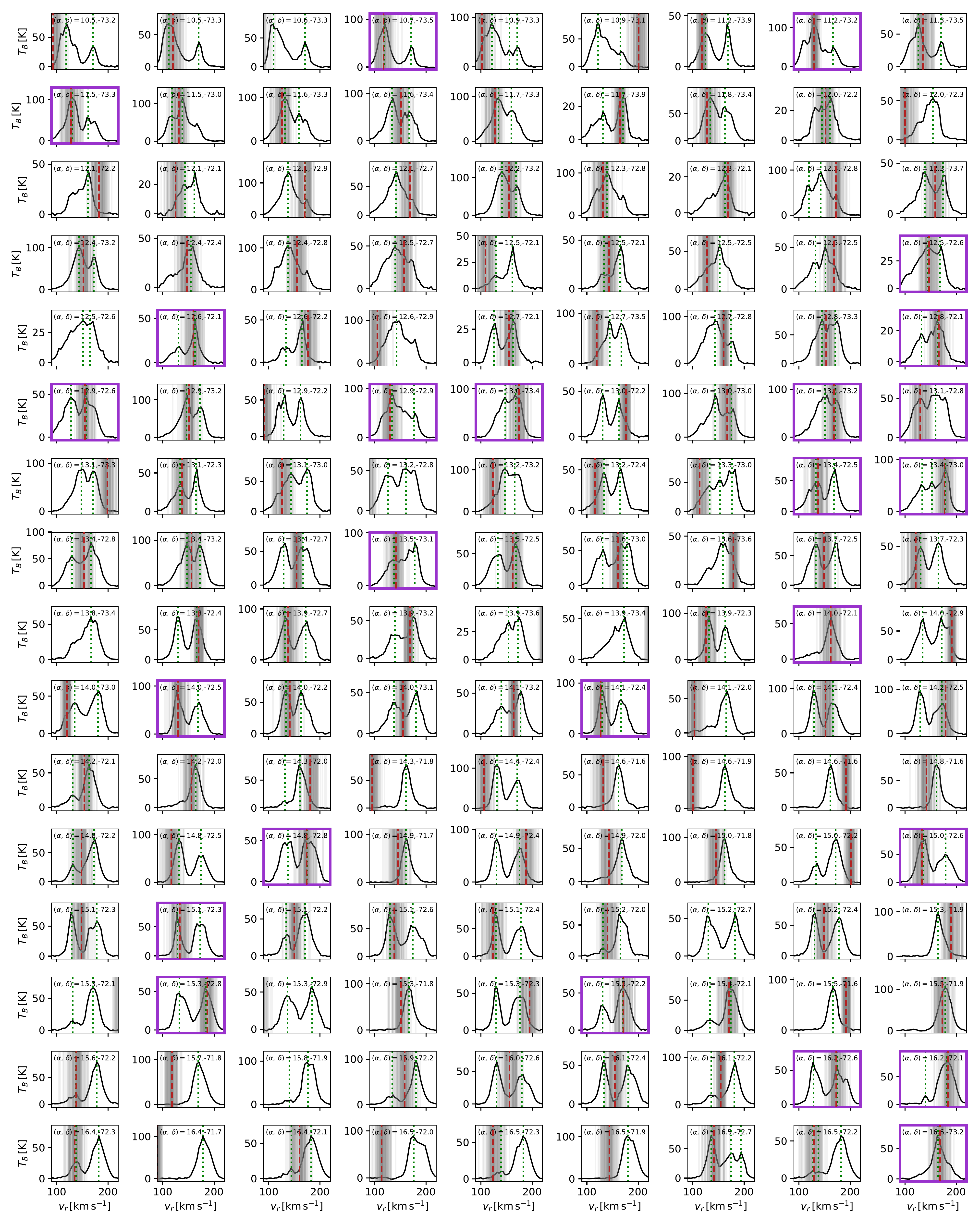}
\caption{Each panel compares the $21\rm\,cm$ $T_B(v_r)$ spectrum and identified \hi\ components ($v_{r,\rm HI}$; dotted green) with the observed stellar radial velocity ($v_{r,\rm star}$; red dashed). The uncertainties in $v_{r,\rm star}$ are illustrated by 50 random draws from a normal distribution with $\mu=v_{r,\rm star}$ and $\sigma=\sigma_{v_{r,\rm star}}$ (grey). The \nstars\ matches (see text) are highlighted by bold, purple plot frames (also see Figure~\ref{f:ob_matched}). If $v_{r,\rm star}$ is not visible within the panel, it falls outside the plotted velocity range. } 
\label{f:match_summary0}
\end{center}
\end{figure*}

\begin{figure*}
\begin{center}
\includegraphics[width=\textwidth]{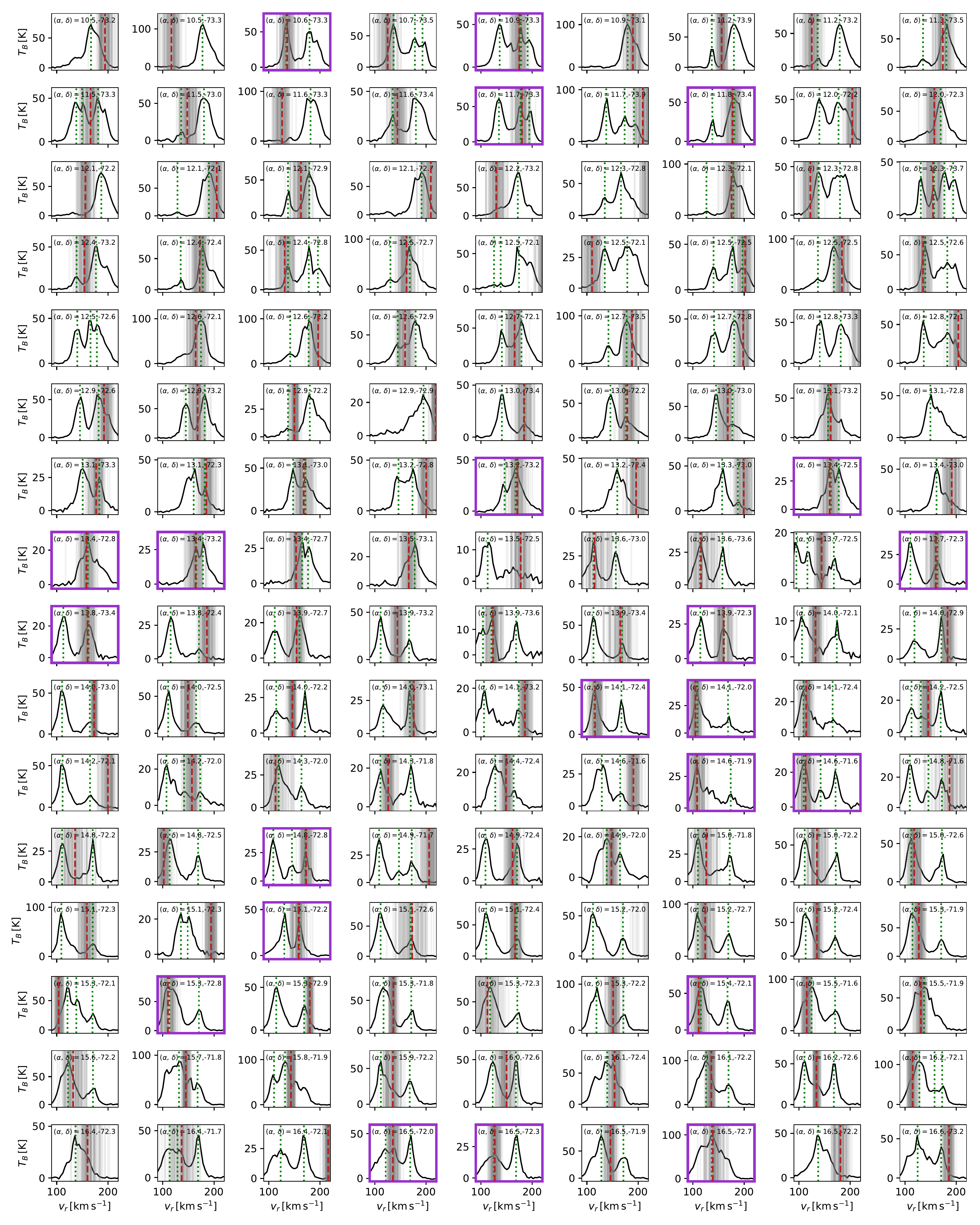}
\caption{See Figure~\ref{f:match_summary0}.}
\label{f:match_summary1}
\end{center}
\end{figure*}

\begin{figure*}
\begin{center}
\includegraphics[width=\textwidth]{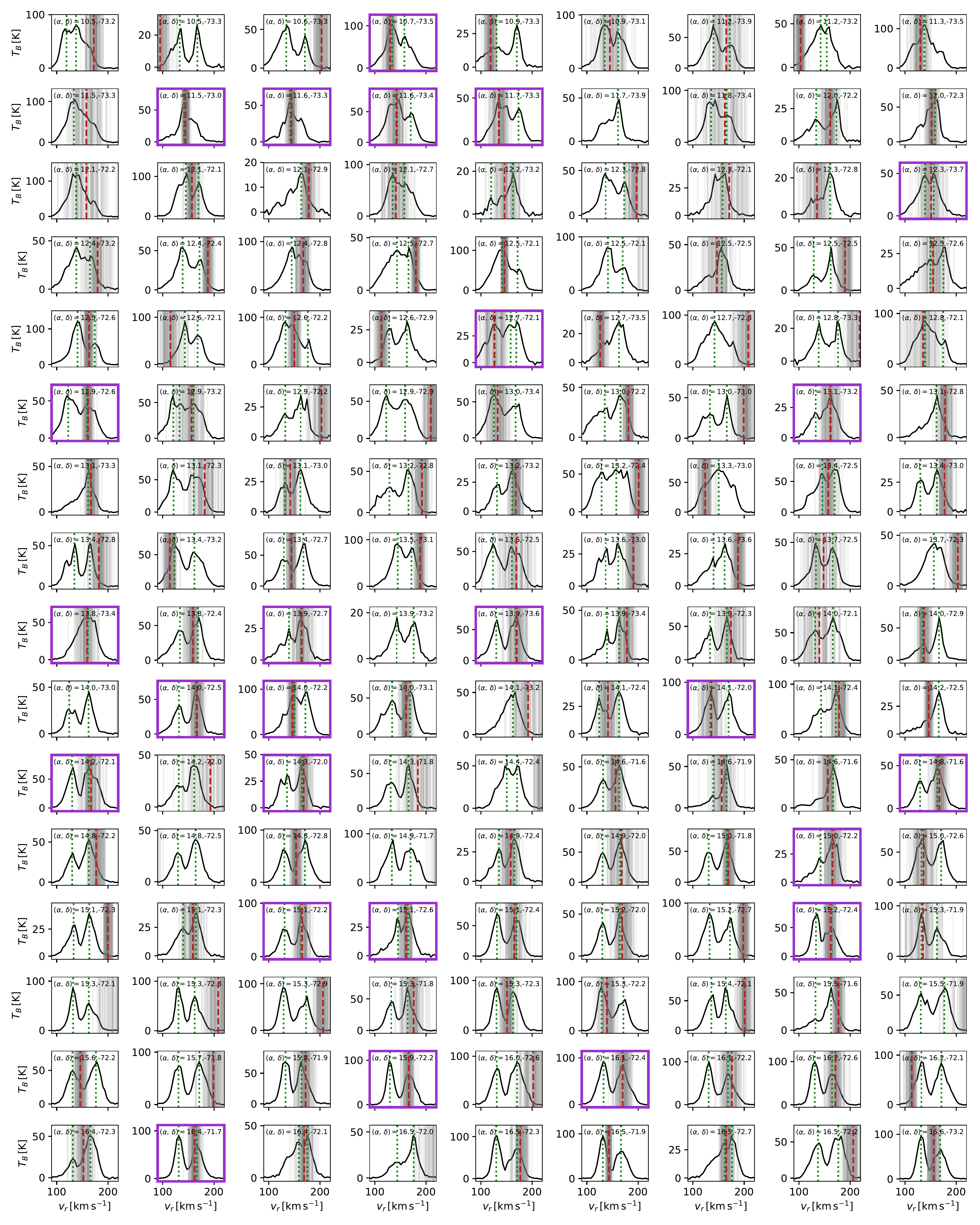}
\caption{See Figure~\ref{f:match_summary0}.}
\label{f:match_summary2}
\end{center}
\end{figure*}

\begin{figure*}
\begin{center}
\includegraphics[width=\textwidth]{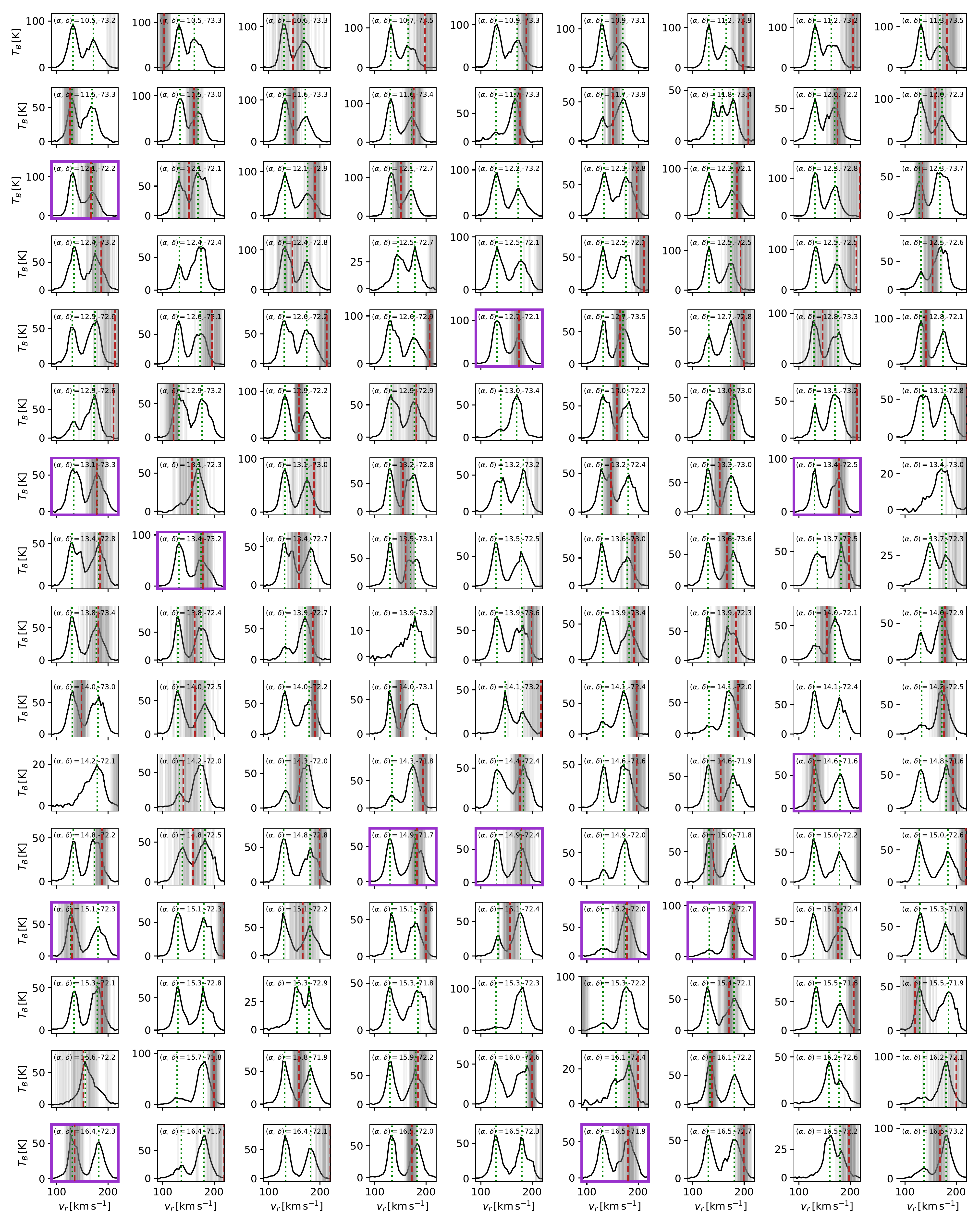}
\caption{See Figure~\ref{f:match_summary0}.}
\label{f:match_summary3}
\end{center}
\end{figure*}

\begin{figure*}
\begin{center}
\includegraphics[width=\textwidth]{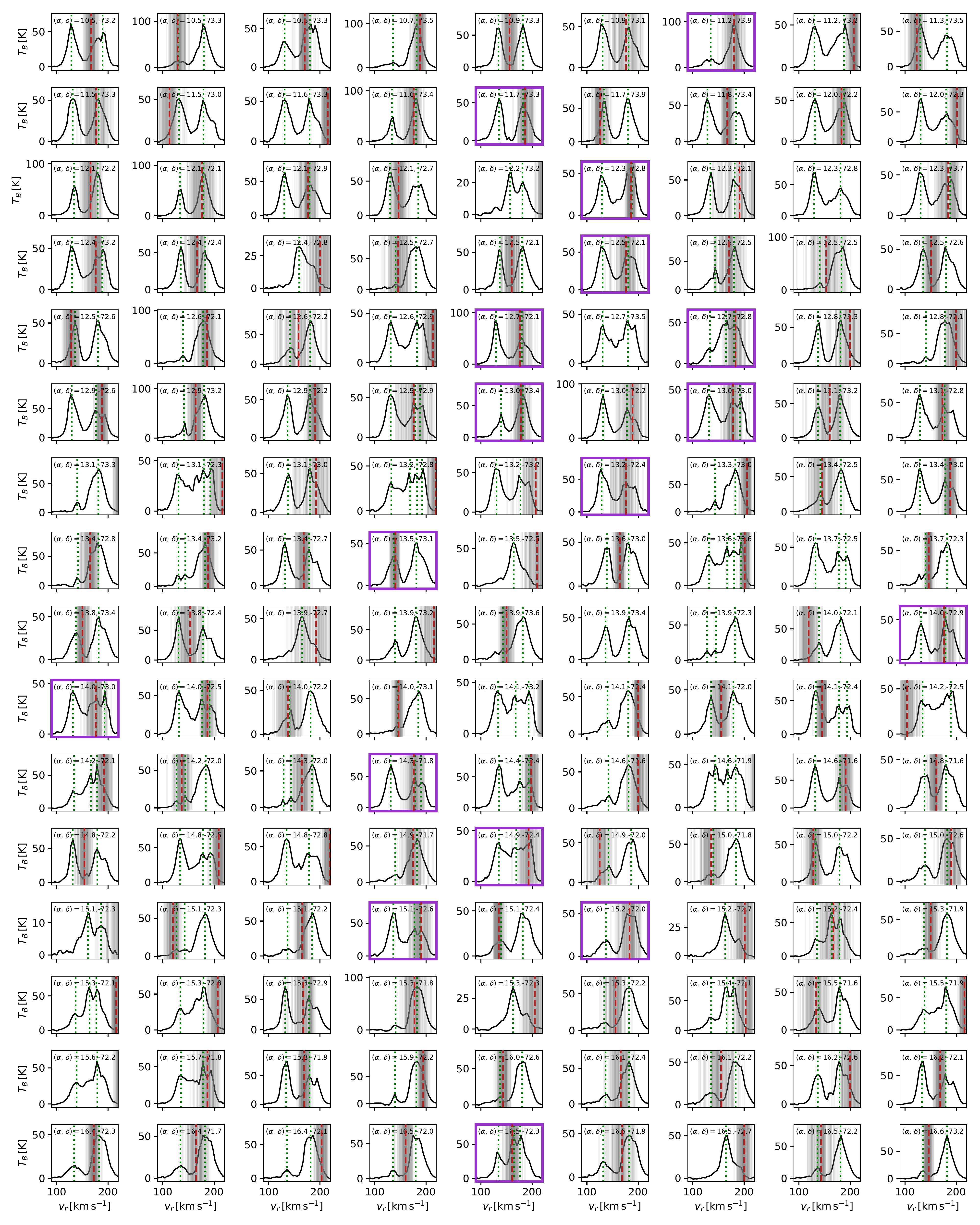}
\caption{See Figure~\ref{f:match_summary0}.}
\label{f:match_summary4}
\end{center}
\end{figure*}

\begin{figure*}
\begin{center}
\includegraphics[width=\textwidth]{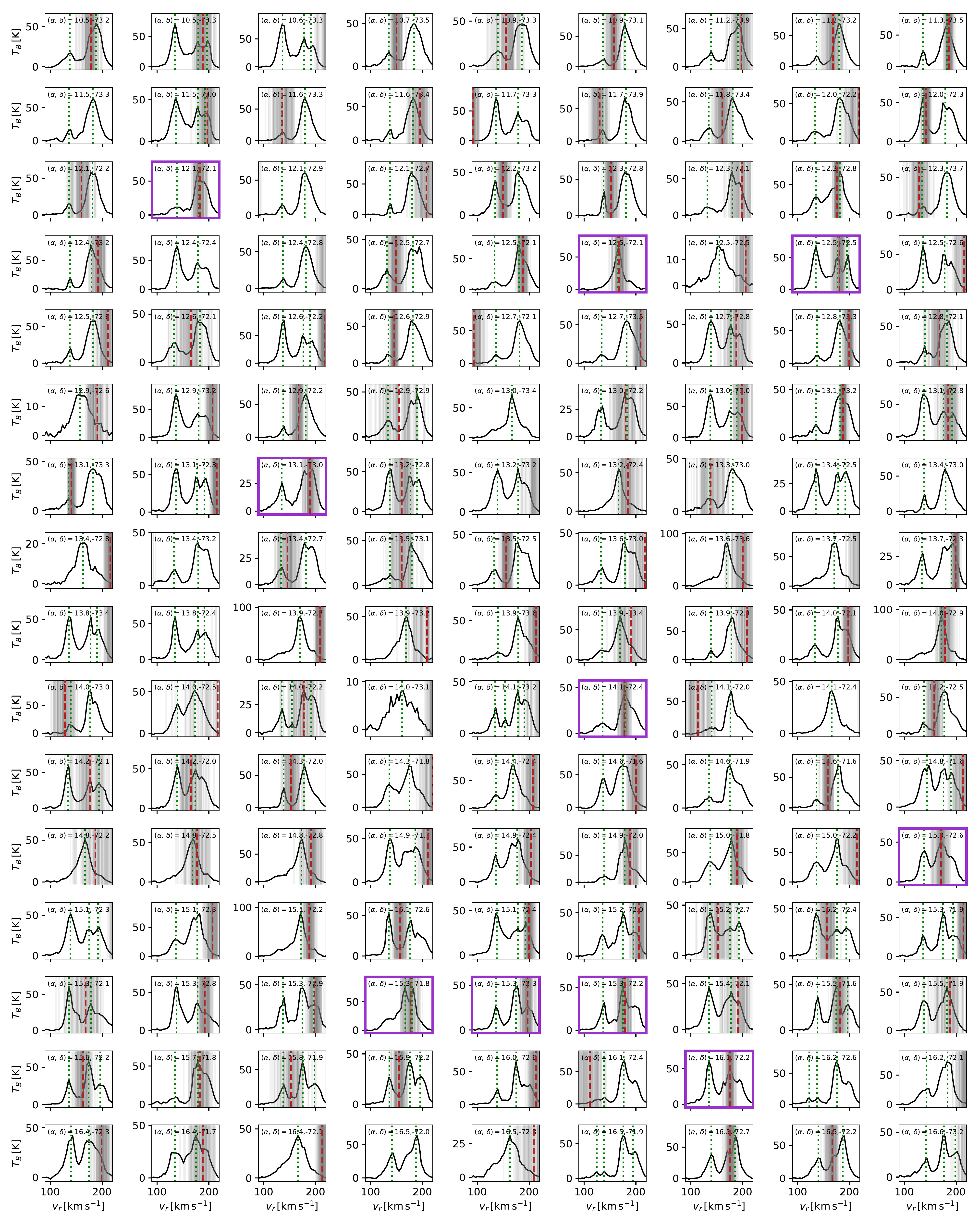}
\caption{See Figure~\ref{f:match_summary0}.}
\label{f:match_summary5}
\end{center}
\end{figure*}

\begin{figure*}
\begin{center}
\includegraphics[width=\textwidth]{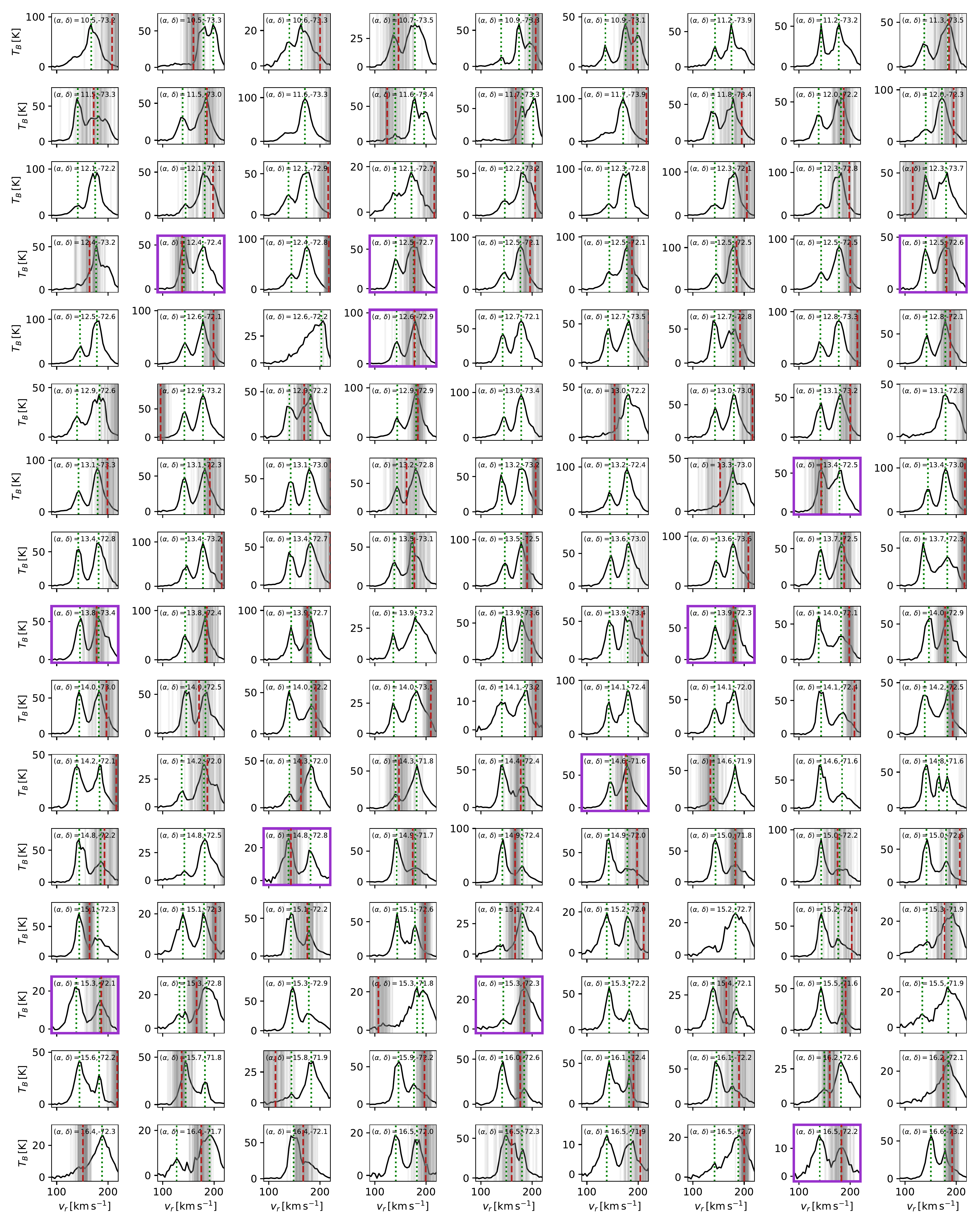}
\caption{See Figure~\ref{f:match_summary0}.}
\label{f:match_summary6}
\end{center}
\end{figure*}

\begin{figure*}
\begin{center}
\includegraphics[width=\textwidth]{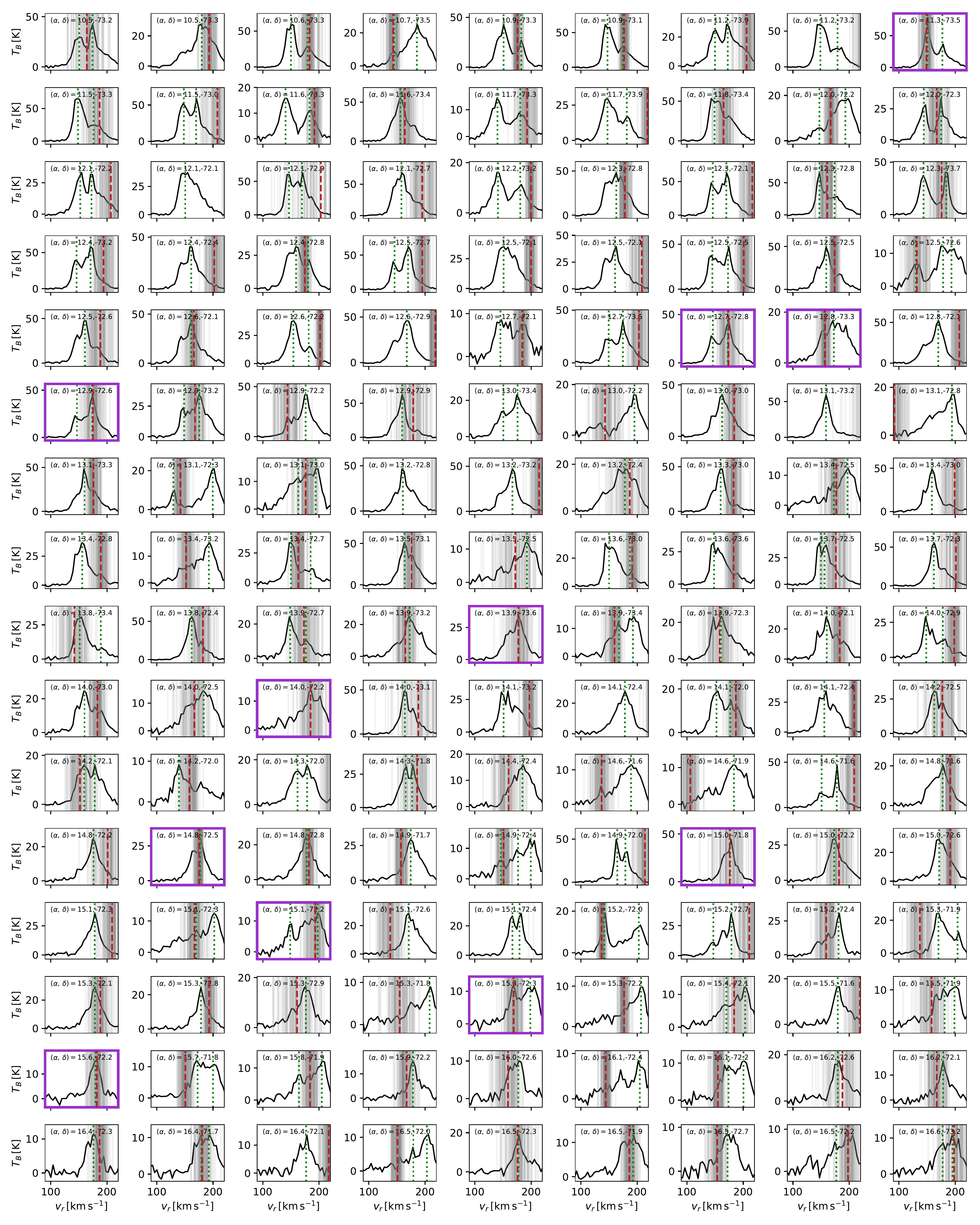}
\caption{See Figure~\ref{f:match_summary0}.}
\label{f:match_summary7}
\end{center}
\end{figure*}

\begin{figure*}
\begin{center}
\includegraphics[width=\textwidth]{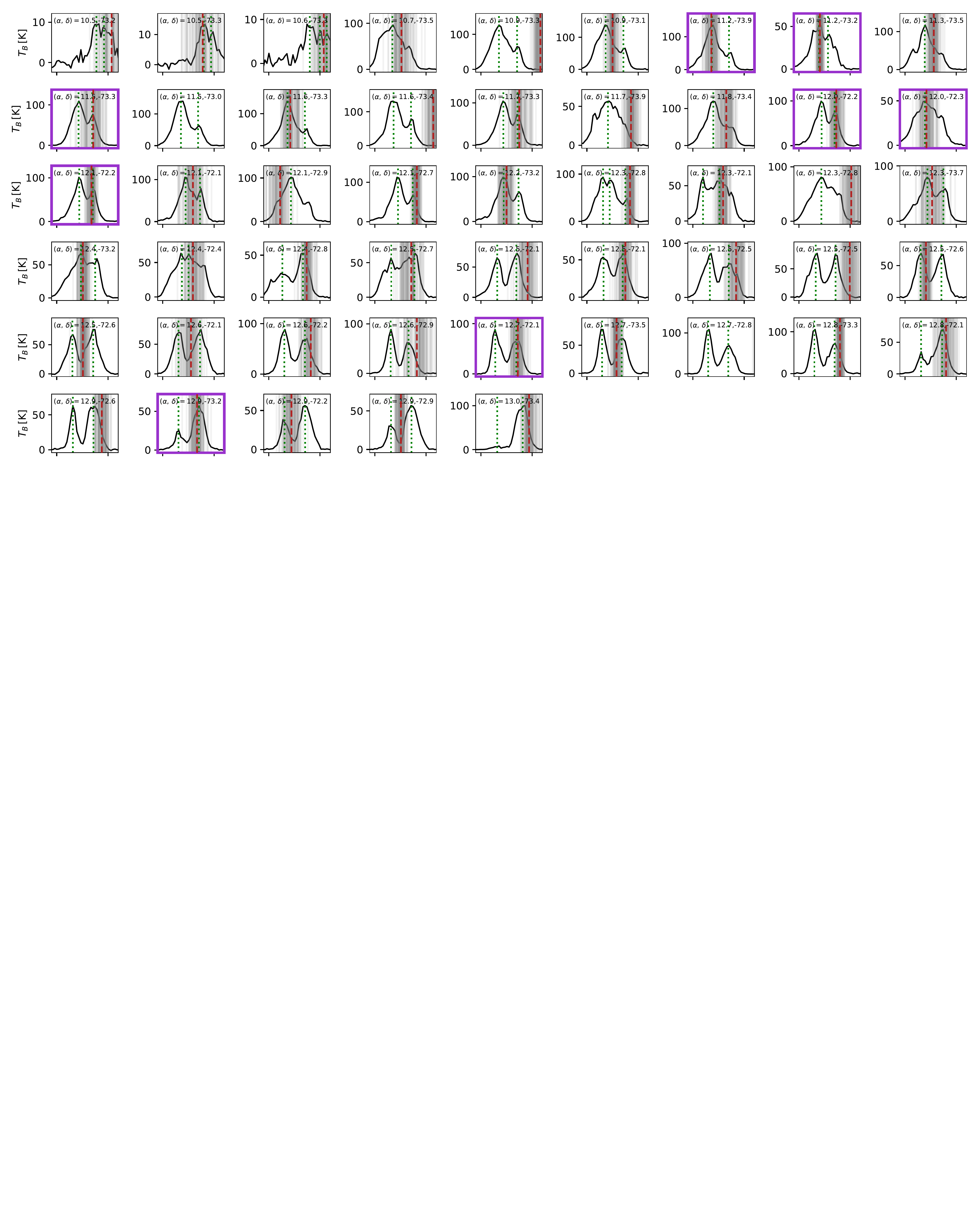}
\vspace{-400pt}
\caption{See Figure~\ref{f:match_summary0}.}
\label{f:match_summary8}
\end{center}
\end{figure*}

In this Appendix, we provide supplemental figures and tables for illustrating the match process between the observed radial velocities of young, massive stars \citep{evans2008, lamb2016}, and the observed radial velocities of significant \hi\ peaks along the same LOS. In Figures~\ref{f:match_summary0} through~\ref{f:match_summary8} we display plots for each of \nstarstot\ stars with significant radial velocities with corresponding proper motion measurements from \emph{Gaia} DR2 \citep{gaiacollaboration2018}. We include the \hi\ spectrum ($T_B(v_r)$) and the identified significant components along each LOS. The \nstars\ matched LOS are identified by purple plot frames (and are also shown in Figure~\ref{f:ob_matched}). 

In Table A, we include the coordinates, velocities, and proper motions for the \nstars\ \hi-matched stars. 

\begin{longtable*}{cc|cccc|llll} 
\hline
\colhead{RA}  &  \colhead{Dec} &   \multicolumn{1}{|c}{$v_{r,\rm star}$} &  \colhead{$\delta{v_{r,\rm star}}$} & \colhead{Type} & \colhead{Ref}  &  \multicolumn{1}{|c}{$\mu_W$} &  \colhead{$\mu_N$} & \colhead{$\delta{\mu_W}$} & \colhead{$\delta{\mu_N}$}  \\[-2mm] 
\multicolumn{2}{c}{(deg)}  & \multicolumn{2}{|c}{($\rm km\,s^{-1}$)} & \colhead{} & \colhead{} & \multicolumn{4}{|c}{($\rm mas\,yr^{-1}$)}  \\[-2mm]
\colhead{(1)} & \colhead{(2)} &  \multicolumn{1}{|c}{(3} & \colhead{(4)} & \colhead{(5)} & \colhead{(6)} &  \multicolumn{1}{|c}{(7)} & \colhead{(8)} & \colhead{(9)} & \colhead{(10)} \\
\hline
7.798  &  -73.993  &  116.0  &  16.0  &  B1-3 (V)  &  2  &  -0.656  &  -1.550  &  0.127  &  0.107    \\   
8.410  &  -73.973  &  160.0  &  9.0  &  B1-3 (IV)  &  2  &  -0.772  &  -1.580  &  0.125  &  0.092    \\   
8.412  &  -74.064  &  160.0  &  9.0  &  B9 (Ib)  &  2  &  -0.501  &  -1.274  &  0.067  &  0.050    \\   
8.565  &  -73.694  &  124.0  &  9.0  &  B3-5 (III)  &  2  &  -0.834  &  -1.402  &  0.136  &  0.107    \\   
8.793  &  -74.122  &  160.0  &  9.0  &  B9 (II)  &  2  &  -0.450  &  -1.190  &  0.082  &  0.069    \\   
9.178  &  -73.137  &  116.0  &  7.0  &  B8 (II)  &  2  &  -0.628  &  -1.088  &  0.056  &  0.049    \\   
9.178  &  -73.794  &  105.0  &  6.0  &  B8 (II)  &  2  &  -0.782  &  -1.291  &  0.097  &  0.069    \\   
9.504  &  -73.652  &  108.0  &  10.0  &  B9 (Ib)  &  2  &  -0.593  &  -1.219  &  0.047  &  0.043    \\   
9.536  &  -73.516  &  113.0  &  9.0  &  B9 (II)  &  2  &  -0.712  &  -1.269  &  0.055  &  0.051    \\   
9.641  &  -73.579  &  173.0  &  7.0  &  B1-3 (II)  &  2  &  -0.418  &  -1.128  &  0.068  &  0.056    \\   
10.452  &  -74.245  &  158.0  &  9.0  &  B0.5 (V)  &  2  &  -1.498  &  -1.849  &  0.191  &  0.154    \\   
10.793  &  -73.575  &  117.0  &  10.0  &  B1.5e$_{3+}$  &  1  &  -0.619  &  -1.062  &  0.076  &  0.065    \\   
10.826  &  -73.446  &  110.0  &  9.0  &  B5 (III)  &  2  &  -0.554  &  -1.242  &  0.061  &  0.058    \\   
11.258  &  -73.234  &  130.0  &  5.0  &  B0 V  &  1  &  -0.692  &  -1.213  &  0.066  &  0.067    \\   
11.393  &  -73.773  &  135.0  &  7.0  &  B8 (II)  &  2  &  -0.596  &  -1.139  &  0.068  &  0.056    \\   
11.419  &  -73.965  &  127.0  &  7.0  &  B1-2 (II)  &  2  &  -0.668  &  -1.371  &  0.102  &  0.074    \\   
11.429  &  -73.163  &  139.0  &  11.0  &  B9 (II)  &  2  &  -0.673  &  -1.007  &  0.067  &  0.059    \\   
11.594  &  -73.388  &  128.0  &  10.0  &  O9.5-B0 III  &  1  &  -1.303  &  -2.176  &  0.200  &  0.174    \\   
11.732  &  -73.332  &  130.0  &  7.0  &  B9 (Ib)  &  2  &  -0.846  &  -1.053  &  0.049  &  0.043    \\   
11.981  &  -73.020  &  136.0  &  8.0  &  B8 (II)  &  2  &  -0.583  &  -1.048  &  0.074  &  0.047    \\   
12.002  &  -72.570  &  144.0  &  4.0  &  B1-2 (III)  &  2  &  -0.606  &  -1.244  &  0.078  &  0.061    \\   
12.014  &  -72.500  &  144.0  &  5.0  &  B9 (Ib)  &  2  &  -0.547  &  -1.209  &  0.050  &  0.040    \\   
12.039  &  -72.730  &  141.0  &  4.0  &  B1-2 (II)  &  2  &  -0.737  &  -1.147  &  0.057  &  0.047    \\   
12.050  &  -73.480  &  142.0  &  10.0  &  B0-5 (II)  &  2  &  -0.798  &  -1.243  &  0.068  &  0.065    \\   
12.060  &  -73.657  &  135.0  &  9.0  &  B2.5 (Ib)  &  2  &  -0.552  &  -1.152  &  0.055  &  0.046    \\   
12.159  &  -73.232  &  171.0  &  13.0  &  B1-5 (II)  &  2  &  -0.770  &  -1.265  &  0.074  &  0.056    \\   
12.495  &  -73.345  &  173.0  &  7.0  &  B0 (IV)  &  2  &  -0.486  &  -1.271  &  0.074  &  0.063    \\   
12.503  &  -72.747  &  142.0  &  10.0  &  B1-5 (III)  &  2  &  -0.731  &  -1.032  &  0.073  &  0.062    \\   
12.520  &  -73.351  &  168.0  &  6.0  &  B0.5 (V)  &  2  &  -0.620  &  -1.026  &  0.078  &  0.069    \\   
12.580  &  -72.655  &  147.0  &  10.0  &  B0 V  &  1  &  -0.506  &  -1.214  &  0.090  &  0.066    \\   
12.607  &  -72.134  &  160.0  &  10.0  &  O9.5 V  &  1  &  -0.778  &  -1.330  &  0.086  &  0.064    \\   
12.846  &  -72.122  &  166.0  &  10.0  &  B1 I  &  1  &  -0.941  &  -1.211  &  0.101  &  0.064    \\   
12.875  &  -72.599  &  126.0  &  16.0  &  B1-5 (III)  &  2  &  -0.562  &  -1.046  &  0.073  &  0.061    \\   
12.927  &  -72.698  &  161.0  &  6.0  &  B1-3 (II)  &  2  &  -1.009  &  -0.897  &  0.198  &  0.143    \\   
12.931  &  -72.624  &  155.0  &  10.0  &  B1 II  &  1  &  -0.565  &  -1.190  &  0.055  &  0.044    \\   
12.999  &  -72.923  &  130.0  &  10.0  &  Be$_3$  &  1  &  -0.867  &  -1.119  &  0.066  &  0.047    \\   
13.035  &  -72.134  &  162.0  &  12.0  &  B0 (II)  &  2  &  -0.734  &  -1.201  &  0.068  &  0.051    \\   
13.129  &  -73.227  &  168.0  &  10.0  &  O9-B0 pe$_{3+}$  &  1  &  -0.747  &  -1.087  &  0.060  &  0.040    \\   
13.148  &  -72.847  &  130.0  &  10.0  &  B1.5e$_{3+}$  &  1  &  -0.755  &  -1.178  &  0.060  &  0.044    \\   
13.267  &  -73.666  &  159.0  &  10.0  &  B5 (II)  &  2  &  -0.405  &  -0.934  &  0.063  &  0.055    \\   
13.304  &  -72.151  &  164.0  &  10.0  &  B0 (V)  &  2  &  -0.818  &  -1.255  &  0.092  &  0.067    \\   
13.361  &  -72.641  &  170.0  &  11.0  &  B0 (III)  &  2  &  -0.727  &  -1.263  &  0.083  &  0.051    \\   
13.415  &  -73.094  &  177.0  &  10.0  &  B0.5 III  &  1  &  -0.538  &  -1.142  &  0.103  &  0.058    \\   
13.434  &  -72.416  &  167.0  &  7.0  &  B1-5 (II)  &  2  &  -0.691  &  -1.304  &  0.070  &  0.062    \\   
13.528  &  -72.667  &  136.0  &  13.0  &  B1-5 (III)  &  2  &  -0.697  &  -1.264  &  0.088  &  0.071    \\   
13.584  &  -72.219  &  168.0  &  8.0  &  B0.5 (IV)  &  2  &  -0.911  &  -1.145  &  0.090  &  0.070    \\   
13.671  &  -72.344  &  167.0  &  9.0  &  B1-5 (II)  &  2  &  -0.750  &  -1.351  &  0.060  &  0.056    \\   
13.745  &  -72.419  &  169.0  &  6.0  &  B5 (II)  &  2  &  -0.696  &  -1.182  &  0.059  &  0.055    \\   
13.748  &  -72.146  &  165.0  &  10.0  &  B8 (II)e?  &  2  &  -0.666  &  -1.247  &  0.080  &  0.065    \\   
13.790  &  -72.465  &  165.0  &  10.0  &  B1-3 (II)  &  2  &  -1.367  &  -1.143  &  0.094  &  0.077    \\   
13.840  &  -72.296  &  162.0  &  7.0  &  B3 (II)  &  2  &  -0.810  &  -1.312  &  0.058  &  0.054    \\   
14.007  &  -72.140  &  162.0  &  10.0  &  O8 V  &  1  &  -0.879  &  -1.391  &  0.081  &  0.064    \\   
14.058  &  -72.500  &  130.0  &  10.0  &  B0.5e$_1$  &  1  &  -0.825  &  -1.283  &  0.048  &  0.047    \\   
14.080  &  -72.515  &  172.0  &  9.0  &  B1-5 (II)  &  2  &  -0.770  &  -1.208  &  0.062  &  0.060    \\   
14.135  &  -72.464  &  127.0  &  10.0  &  Be$_{4+}$  &  1  &  -0.860  &  -1.295  &  0.070  &  0.068    \\   
14.200  &  -72.479  &  166.0  &  6.0  &  B1-3 (II)  &  2  &  -0.672  &  -1.231  &  0.059  &  0.058    \\   
14.206  &  -72.755  &  170.0  &  10.0  &  B0.5 (IV)  &  2  &  -0.693  &  -1.261  &  0.065  &  0.057    \\   
14.335  &  -73.590  &  165.0  &  10.0  &  B0-5 (II)  &  2  &  -0.557  &  -1.232  &  0.076  &  0.062    \\   
14.406  &  -72.292  &  126.0  &  7.0  &  B8 (II)  &  2  &  -1.132  &  -1.398  &  0.096  &  0.072    \\   
14.464  &  -72.508  &  167.0  &  9.0  &  B2 (II)  &  2  &  -0.765  &  -1.237  &  0.065  &  0.053    \\   
14.732  &  -72.535  &  174.0  &  6.0  &  B0 (Ib)  &  2  &  -0.559  &  -1.169  &  0.064  &  0.057    \\   
14.864  &  -72.810  &  174.0  &  10.0  &  O8.5 V  &  1  &  -0.736  &  -1.268  &  0.086  &  0.064    \\   
14.941  &  -72.749  &  178.0  &  9.0  &  B0 (Iab)  &  2  &  -0.688  &  -1.316  &  0.065  &  0.055    \\   
14.979  &  -72.218  &  167.0  &  9.0  &  B9 (Ib)  &  2  &  -0.881  &  -1.227  &  0.054  &  0.039    \\   
15.071  &  -72.397  &  178.0  &  7.0  &  B0.5 (V)  &  2  &  -0.818  &  -1.189  &  0.076  &  0.059    \\   
15.085  &  -72.690  &  133.0  &  10.0  &  B0 III  &  1  &  -0.703  &  -1.509  &  0.060  &  0.051    \\   
15.116  &  -72.648  &  178.0  &  9.0  &  B0-5 (III)  &  2  &  -0.701  &  -1.156  &  0.105  &  0.073    \\   
15.358  &  -72.885  &  187.0  &  10.0  &  B1.5 III  &  1  &  -0.608  &  -1.304  &  0.070  &  0.051    \\   
15.387  &  -72.279  &  171.0  &  10.0  &  O6 V  &  1  &  -0.996  &  -1.074  &  0.094  &  0.065    \\   
15.479  &  -72.124  &  170.0  &  10.0  &  O7 III((f))e$_1$  &  1  &  -0.825  &  -1.143  &  0.080  &  0.069    \\   
15.504  &  -72.453  &  130.0  &  8.0  &  O8.5 III((f))  &  2  &  -0.961  &  -1.280  &  0.077  &  0.060    \\   
15.541  &  -72.584  &  182.0  &  8.0  &  O9.5 III  &  2  &  -0.822  &  -1.213  &  0.079  &  0.071    \\   
15.548  &  -72.457  &  179.0  &  14.0  &  B0.5 (V)  &  2  &  -0.947  &  -1.361  &  0.082  &  0.062    \\   
15.662  &  -72.668  &  130.0  &  12.0  &  B1-5 (II)  &  2  &  -0.659  &  -1.240  &  0.077  &  0.063    \\   
15.693  &  -72.153  &  178.0  &  12.0  &  B2 (IV)  &  2  &  -0.851  &  -1.266  &  0.106  &  0.077    \\   
15.699  &  -72.122  &  180.0  &  4.0  &  B8 (Ib)  &  2  &  -0.856  &  -1.249  &  0.055  &  0.041    \\   
15.884  &  -72.654  &  180.0  &  10.0  &  B1-5 (III)  &  2  &  -0.794  &  -1.257  &  0.101  &  0.087    \\   
15.953  &  -72.207  &  180.0  &  10.0  &  B1-3 (III)  &  2  &  -0.970  &  -1.186  &  0.079  &  0.064    \\   
16.046  &  -72.428  &  186.0  &  9.0  &  B0-5 (III)  &  2  &  -1.024  &  -1.251  &  0.092  &  0.073    \\   
16.063  &  -72.693  &  176.0  &  9.0  &  B0-5 (II)  &  2  &  -0.654  &  -1.256  &  0.078  &  0.057    \\   
16.097  &  -72.807  &  186.0  &  5.0  &  B1-3 (III)  &  2  &  -0.741  &  -1.141  &  0.061  &  0.053    \\   
16.222  &  -72.676  &  173.0  &  10.0  &  O8.5 V  &  1  &  -0.676  &  -1.121  &  0.074  &  0.064    \\   
16.227  &  -72.608  &  176.0  &  12.0  &  B1-5 (III)  &  2  &  -0.866  &  -1.378  &  0.118  &  0.092    \\   
16.240  &  -73.071  &  183.0  &  12.0  &  B9 (Ib)  &  2  &  -0.887  &  -1.126  &  0.055  &  0.042    \\   
16.274  &  -72.135  &  184.0  &  10.0  &  O9.5 V  &  1  &  -0.704  &  -1.257  &  0.087  &  0.054    \\   
16.276  &  -72.300  &  178.0  &  7.0  &  B1-3 (III)  &  2  &  -0.922  &  -1.146  &  0.109  &  0.060    \\   
16.281  &  -72.805  &  178.0  &  9.0  &  O5 V((f))  &  2  &  -0.751  &  -1.321  &  0.072  &  0.065    \\   
16.320  &  -72.670  &  176.0  &  15.0  &  B0-3 (V)  &  2  &  -0.684  &  -1.229  &  0.108  &  0.089    \\   
16.364  &  -72.478  &  140.0  &  5.0  &  B1-5 (Ib)  &  2  &  -0.966  &  -1.248  &  0.081  &  0.056    \\   
16.419  &  -72.577  &  176.0  &  13.0  &  B1-3 (III)  &  2  &  -0.799  &  -1.285  &  0.083  &  0.056    \\   
16.431  &  -72.801  &  176.0  &  12.0  &  B1-5 (II)  &  2  &  -0.770  &  -1.150  &  0.071  &  0.063    \\   
16.490  &  -72.663  &  176.0  &  10.0  &  B2 (IV)  &  2  &  -0.941  &  -1.209  &  0.093  &  0.066    \\   
16.543  &  -72.840  &  193.0  &  12.0  &  B3 (II)  &  2  &  -0.731  &  -1.244  &  0.063  &  0.053    \\   
16.600  &  -72.708  &  190.0  &  11.0  &  B2 (V)  &  2  &  -1.132  &  -1.433  &  0.137  &  0.098    \\   
16.612  &  -72.435  &  183.0  &  12.0  &  B1-3 (III)  &  2  &  -0.967  &  -1.112  &  0.093  &  0.069    \\   
16.637  &  -73.262  &  168.0  &  10.0  &  B0e$_3$  &  1  &  -0.825  &  -1.361  &  0.124  &  0.076    \\   
16.708  &  -73.080  &  161.0  &  9.0  &  B5 (III)  &  2  &  -0.999  &  -1.257  &  0.083  &  0.058    \\   
16.713  &  -72.556  &  136.0  &  10.0  &  O8.5 V  &  1  &  -1.093  &  -1.269  &  0.087  &  0.069    \\   
16.757  &  -72.799  &  175.0  &  10.0  &  B1.5e$_2$  &  1  &  -0.669  &  -1.316  &  0.065  &  0.052    \\   
16.916  &  -72.430  &  183.0  &  8.0  &  B0 (V)  &  2  &  -0.949  &  -1.231  &  0.110  &  0.077    \\   
16.985  &  -73.516  &  168.0  &  8.0  &  B0-5 (V)  &  2  &  -0.777  &  -1.354  &  0.128  &  0.098    \\   
17.030  &  -72.693  &  180.0  &  10.0  &  B1.5e$_3$  &  1  &  -0.718  &  -1.264  &  0.056  &  0.050    \\   
17.133  &  -72.240  &  177.0  &  10.0  &  O9.5 III  &  1  &  -0.786  &  -1.230  &  0.114  &  0.071    \\   
17.191  &  -72.523  &  189.0  &  6.0  &  B0-5 (II)  &  2  &  -1.055  &  -1.368  &  0.084  &  0.073    \\   
17.362  &  -72.384  &  178.0  &  6.0  &  B1-5 (II)e  &  2  &  -0.785  &  -1.336  &  0.078  &  0.064    \\   
17.451  &  -72.505  &  155.0  &  10.0  &  O9 V  &  1  &  -1.243  &  -1.336  &  0.096  &  0.085    \\   
17.561  &  -72.870  &  171.0  &  9.0  &  B1-3 (IV)  &  2  &  -1.122  &  -1.528  &  0.078  &  0.082    \\   
17.800  &  -72.369  &  198.0  &  6.0  &  O9.5 III  &  2  &  -0.846  &  -1.355  &  0.084  &  0.057    \\   
17.803  &  -73.159  &  177.0  &  9.0  &  B1-3 (III)  &  2  &  -0.838  &  -1.241  &  0.092  &  0.065    \\   
17.832  &  -72.475  &  197.0  &  8.0  &  B2.5 (II)  &  2  &  -0.836  &  -0.919  &  0.067  &  0.061    \\   
17.833  &  -72.366  &  179.0  &  8.0  &  O8 III  &  2  &  -0.933  &  -1.323  &  0.124  &  0.072    \\   
17.998  &  -72.743  &  176.0  &  5.0  &  B1-5 (III)  &  2  &  -1.248  &  -1.464  &  0.097  &  0.090    \\   
18.746  &  -72.872  &  138.0  &  11.0  &  B0-5 (IV)  &  2  &  -1.022  &  -1.257  &  0.102  &  0.093    \\   
18.782  &  -73.411  &  177.0  &  9.0  &  B1-5 (II)  &  2  &  -0.871  &  -1.243  &  0.084  &  0.064    \\   
18.864  &  -73.426  &  181.0  &  10.0  &  B2.5 (II)  &  2  &  -1.145  &  -1.209  &  0.084  &  0.061    \\   
18.886  &  -73.318  &  177.0  &  15.0  &  B1-5 (III)  &  2  &  -0.852  &  -1.358  &  0.126  &  0.074    \\   
18.958  &  -73.107  &  184.0  &  8.0  &  B3-5 (III)  &  2  &  -0.849  &  -1.279  &  0.114  &  0.091    \\   
19.055  &  -72.807  &  143.0  &  12.0  &  B1-5 (III)  &  2  &  -1.039  &  -1.518  &  0.108  &  0.093    \\   
19.320  &  -73.205  &  178.0  &  12.0  &  B1-3 (III)  &  2  &  -0.859  &  -1.422  &  0.096  &  0.069    \\   
19.433  &  -73.056  &  179.0  &  6.0  &  B0-5 (III)  &  2  &  -0.548  &  -1.433  &  0.097  &  0.067    \\   
19.771  &  -72.845  &  176.0  &  13.0  &  B9: (II)  &  2  &  -1.134  &  -1.355  &  0.096  &  0.074    \\   
20.209  &  -73.692  &  187.0  &  14.0  &  B3-5 (III)e  &  2  &  -0.854  &  -1.290  &  0.083  &  0.064    \\   
20.258  &  -72.727  &  184.0  &  7.0  &  B0.5 (V)  &  2  &  -1.121  &  -1.470  &  0.103  &  0.074    \\   
20.647  &  -74.035  &  183.0  &  12.0  &  B0-5 (IV)  &  2  &  -0.767  &  -1.337  &  0.128  &  0.097    \\   
20.813  &  -72.992  &  149.0  &  7.0  &  B1-5 (III)  &  2  &  -0.834  &  -1.314  &  0.112  &  0.094    \\   
21.213  &  -73.450  &  185.0  &  10.0  &  O8 V  &  1  &  -1.149  &  -1.324  &  0.068  &  0.058    \\   
21.213  &  -73.100  &  168.0  &  10.0  &  B0.2 V  &  1  &  -1.185  &  -1.135  &  0.087  &  0.069    \\   
21.362  &  -74.292  &  184.0  &  7.0  &  B8 (II)  &  2  &  -0.989  &  -1.315  &  0.066  &  0.047    \\   
21.415  &  -73.153  &  174.0  &  10.0  &  B1-5 (III)  &  2  &  -0.919  &  -0.766  &  0.112  &  0.091    \\   
21.449  &  -73.144  &  175.0  &  11.0  &  B9 (II)  &  2  &  -1.212  &  -1.170  &  0.083  &  0.062    \\   
21.647  &  -73.255  &  171.0  &  10.0  &  O9.5 III  &  1  &  -0.948  &  -1.138  &  0.089  &  0.063    \\   
21.990  &  -73.171  &  160.0  &  10.0  &  B0.5 V  &  1  &  -1.070  &  -1.277  &  0.078  &  0.064    \\   
22.132  &  -73.149  &  177.0  &  13.0  &  B0-5 (IV)  &  2  &  -1.047  &  -1.217  &  0.121  &  0.092    \\   
22.236  &  -73.387  &  158.0  &  10.0  &  B0.7 V  &  1  &  -0.884  &  -1.186  &  0.112  &  0.071    \\   
22.278  &  -72.606  &  185.0  &  14.0  &  B0-5 (IV)e  &  2  &  -1.423  &  -1.047  &  0.101  &  0.071    \\   
22.545  &  -73.316  &  164.0  &  10.0  &  B[e]  &  1  &  -0.778  &  -1.139  &  0.066  &  0.043    \\   
22.621  &  -73.327  &  175.0  &  4.0  &  B1-5 (III)  &  2  &  -0.753  &  -1.220  &  0.140  &  0.088    \\   
22.883  &  -72.652  &  167.0  &  11.0  &  B0-5 (IV)  &  2  &  -1.198  &  -1.198  &  0.097  &  0.078    \\   
23.808  &  -73.079  &  169.0  &  11.0  &  B1-5 (III)  &  2  &  -1.248  &  -1.352  &  0.087  &  0.073    \\   
23.980  &  -73.641  &  182.0  &  10.0  &  B1-5 (III)  &  2  &  -1.366  &  -1.141  &  0.123  &  0.093    \\   
24.250  &  -73.336  &  181.0  &  9.0  &  B3-5 (III)  &  2  &  -1.057  &  -1.158  &  0.129  &  0.085    \\   
24.303  &  -73.561  &  176.0  &  16.0  &  B1-5 (III)  &  2  &  -1.347  &  -1.359  &  0.124  &  0.094    \\   
24.382  &  -73.167  &  198.0  &  11.0  &  B1-5 (IV)  &  2  &  -1.187  &  -1.364  &  0.151  &  0.101    \\

\hline   
\caption{Parameters for the \nstars{} stars used in the kinematic comparison. (1-2): RA, Dec coordinates; (3-4): Observed radial velocity and uncertainty; (5) Stellar type; (6) Radial velocity reference, 1=\citet{lamb2016}, 2=\citet{evans2008}; (7-10): Proper motions and uncertainties in the West (7,9) and North (8,10) directions from \citet{gaiacollaboration2018}.}
\end{longtable*}

\end{document}